\documentclass[aps,pra,amsmath,amssymb,reprint,nofootinbib,floatfix,groupedaddress]{revtex4-1}

\usepackage{graphicx}
\usepackage[bookmarks=false,colorlinks=true,linkcolor=blue,filecolor=blue,citecolor=blue,urlcolor=blue]{hyperref}

\usepackage{bm}
\usepackage{soul}

\usepackage{times}
\usepackage[normalem]{ulem}

\begin{document}

\title{Measurement-induced quantum criticality under continuous monitoring}

\author{Yohei Fuji}
\affiliation{Department of Applied Physics, University of Tokyo, Tokyo 113-8656, Japan}

\author{Yuto Ashida}
\affiliation{Department of Applied Physics, University of Tokyo, Tokyo 113-8656, Japan}

\date{\today}

\begin{abstract}
We investigate entanglement phase transitions from volume-law to area-law entanglement in a quantum many-body state under continuous position measurement on the basis of the quantum trajectory approach. 
We find the signatures of the transitions as peak structures in the mutual information as a function of measurement strength, as previously reported for random unitary circuits with projective measurements. 
At the transition points, the entanglement entropy scales logarithmically and various physical quantities scale algebraically, implying emergent conformal criticality, for both integrable and nonintegrable one-dimensional interacting Hamiltonians; however, such transitions have been argued to be absent in noninteracting regimes in some previous studies. 
With the aid of $U(1)$ symmetry in our model, the measurement-induced criticality exhibits a spectral signature
resembling a Tomonaga-Luttinger liquid theory from symmetry-resolved entanglement. 
These intriguing critical phenomena are unique to steady-state regimes of the conditional dynamics at the single-trajectory level, and are absent in the unconditional dynamics obeying the Lindblad master equation, in which the system ends up with the featureless, infinite-temperature mixed state.
We also propose a possible experimental setup to test the predicted entanglement transition based on the subsystem particle-number fluctuations. 
This quantity should readily be measured by the current techniques of quantum gas microscopy and is in practice easier to obtain than the entanglement entropy itself.
\end{abstract}

\maketitle
\tableofcontents

\section{Introduction}
Entanglement is an important diagnostic tool for many-body quantum systems from various aspects, such as critical phenomena, nontrivial topology, and dynamics \cite{Amico08, Eisert10, Laflorencie16}. 
Given a low-entangled state as an initial state, unitary time evolutions generate entanglement between two distant subregions of the system. 
For generic systems exhibiting thermalization, a bipartite entanglement entropy, a typical measure of entanglement, grows linearly in time and saturates to a steady-state value after a long time, whose amount is in general proportional to the volume of the subregion. 
This is known as the \emph{volume law} of the entanglement entropy, and is a characteristic feature of equilibrium states at finite temperatures.

Quantum systems that do not obey thermalizing behaviors have intensively been studied, including many-body localized states \cite{Nandkishore15, Abanin19} and quantum scar states \cite{Turner18a, Turner18b}. 
For example, the many-body localized states are characterized by a bipartite entanglement entropy scaling with the surface area of the subregion, which is called the \emph{area law}. 
Recently, an alternative way to realize non-thermalizing states with area-law entanglement has been proposed by the use of projective measurements. 
As the projection measurement of a local operator disentangles the measured local state from the rest of the system, a frequent operation of the projective measurements will tend to decrease the entanglement entropy. 
This protocol has been investigated for one-dimensional (1D) systems in combination with the unitary time evolution generated by random unitary circuits \cite{Chan19, Skinner19, Li18}, which are considered as simple toy models to capture the entanglement growth in thermalizing systems \cite{Nahum17, Nahum18, vonKeyserlingk18, Khemani18, Rakovszky18}. 
It has been shown that the competition between the random unitary dynamics and projective measurements gives rise to a quantum phase transition in entanglement from the volume law to area law as the measurement strength is increased.
Remarkably, a pure state at this transition point exhibits a logarithmic scaling of the entanglement entropy with the subregion size and power-law behaviors in various correlations, which are prominent features of critical systems described by the (1+1)-D conformal field theory (CFT). 
Such measurement-induced entanglement transitions have been studied in different setups of circuit models and from various physical aspects \cite{Li19, Szyniszewski19, Choi19, Gullans19a, Gullans19b, Zabalo20, Zhang20, Fan20, LopezPiqueres20, Li20, Shtanko20, Levasani20, Sang20, Ippoliti20, Chen20}.
There have also been several attempts to analytically extract universal quantities at the criticality by mapping circuit models to statistical mechanical models of percolation \cite{Skinner19, Bao20, Jian20}.

A natural question is whether or not a measurement-induced criticality (MIC) also occurs in generic open many-body systems that are directly relevant to realistic physical setups. 
For free-fermion systems under continuous measurement, it has been argued that there would be no transitions on the basis of the collapsed quasipartcle pair ansatz, which could also be applied to integrable models \cite{Cao19}; as long as this ansatz is applicable, the entanglement entropy should obey the area law for any finite strength of the measurement. 
On the other hand, a non-integrable, Bose-Hubbard model under random projective measurements has been shown to exhibit a MIC with scaling behaviors expected from those for the random unitary circuits \cite{Tang20}. 
There have also been related studies for the (unstable) Bose-Hubbard model with two-body loss \cite{Goto20} and a quantum Ising model with random projective measurements \cite{Rossini20}. 

In this paper, we study the dynamics of many-body quantum states in a 1D interacting system of bosons under continuous measurement of local particle numbers. 
Following Refs.~\cite{Cao19, Goto20}, we employ the quantum trajectory approach \cite{Daley14} to directly access the entanglement dynamics of pure states conditioned on measurement outcomes, instead of focusing on the unconditional dynamics of density matrices through the Lindblad master equation. 
The latter case, rather, leads to trivial featureless steady states given by the infinite-temperature equilibrium states for generic nonintegrable Hamiltonians. 
In the quantum trajectory approach, measurements occur as quantum jumps associated with the nonunitary time evolution governed by a non-Hermitian Hamiltonian. 
In contrast to previous studies with random projective measurements \cite{Chan19, Skinner19, Li18, Li19, Szyniszewski19, Choi19, Gullans19a, Gullans19b, Zabalo20, Zhang20, Fan20, Li20, Tang20}, the probability distribution of measurement is in general \emph{not} uniform in both space and time and depends on the expectation values of observables under the nonunitary time evolution. 
However, in our case of continuous position measurement, the $U(1)$ symmetry associated with particle-number conservation renders the non-Hermitian part of the Hamiltonian just a constant and sets the probability distribution to be uniform in time, i.e., the frequency of observing quantum jumps remains the same. 
The probability distribution for measured positions of particles also becomes effectively uniform due to quick relaxation of local occupation numbers in the regime of time scale that we are interested in. 
Therefore, our model of continuous measurement in the view of quantum trajectory is considered to give the dynamics close to those of unitary circuit models with random insertions of projective measurements, apart from the fact that the unitary time evolution in our case is given by the Hamiltonian.

From numerical simulations for finite-size systems, we find that the entanglement entropy in steady states averaged over different realizations of quantum trajectories undergoes a phase transition from the volume-law to area-law regimes as the measurement strength is increased.
At this transition point, the results feature various signatures of emergent conformal invariance and thereby the MIC, such as logarithmic scaling of the von Neumann entanglement entropy in subsystem sizes, algebraic decays of correlation functions, and quadratic dispersions of the $U(1)$-symmetry resolved entanglement suggestive of a free boson CFT or a Tomonaga-Luttinger liquid theory; however, universal quantities at the MIC do not resemble those for the conventional quantum criticality described by the (1+1)-D CFT.
As proposed in Ref.~\cite{Li19} for unitary circuits, the transition point can be conveniently located by a peak structure of the mutual information or certain correlation functions. 
Importantly, these characteristic behaviors in the MIC appear in both integrable and nonintegrable regimes of the model, suggesting that the entanglement dynamics in integrable systems might be more intricate than that captured by a simple quasiparticle picture \cite{Cao19}. 
We emphasize that our model is directly relevant to the current technology of ultracold atomic experiments. 
In particular, to test our theoretical results, we propose probing several physical observables such as the subsystem particle-number fluctuations, which should be measured by the site-resolved detection techniques enabled by quantum gas microscopy.

This paper is organized as follows.
In Sec.~\ref{sec:Model}, we introduce our model and the quantum trajectory approach for continuous measurement along with its simulation protocol. 
In Sec.~\ref{sec:Results}, we present our main results of numerical simulation and show characteristic behaviors of the entanglement and correlation functions at the MIC. 
In Sec.~\ref{sec:ApplExp}, we discuss applications of our model to ultracold atomic experiments and propose useful probes to detect the MIC. 
We conclude our paper in Sec.~\ref{sec:Conclusion}. 
Appendix~\ref{app:Scaling} provides a detailed comparison of different scaling analyses for the entanglement entropy in the vicinities of MICs. 

\section{Model}
\label{sec:Model}

\subsection{Hamiltonian}

\begin{figure}
\includegraphics[clip,width=0.38\textwidth]{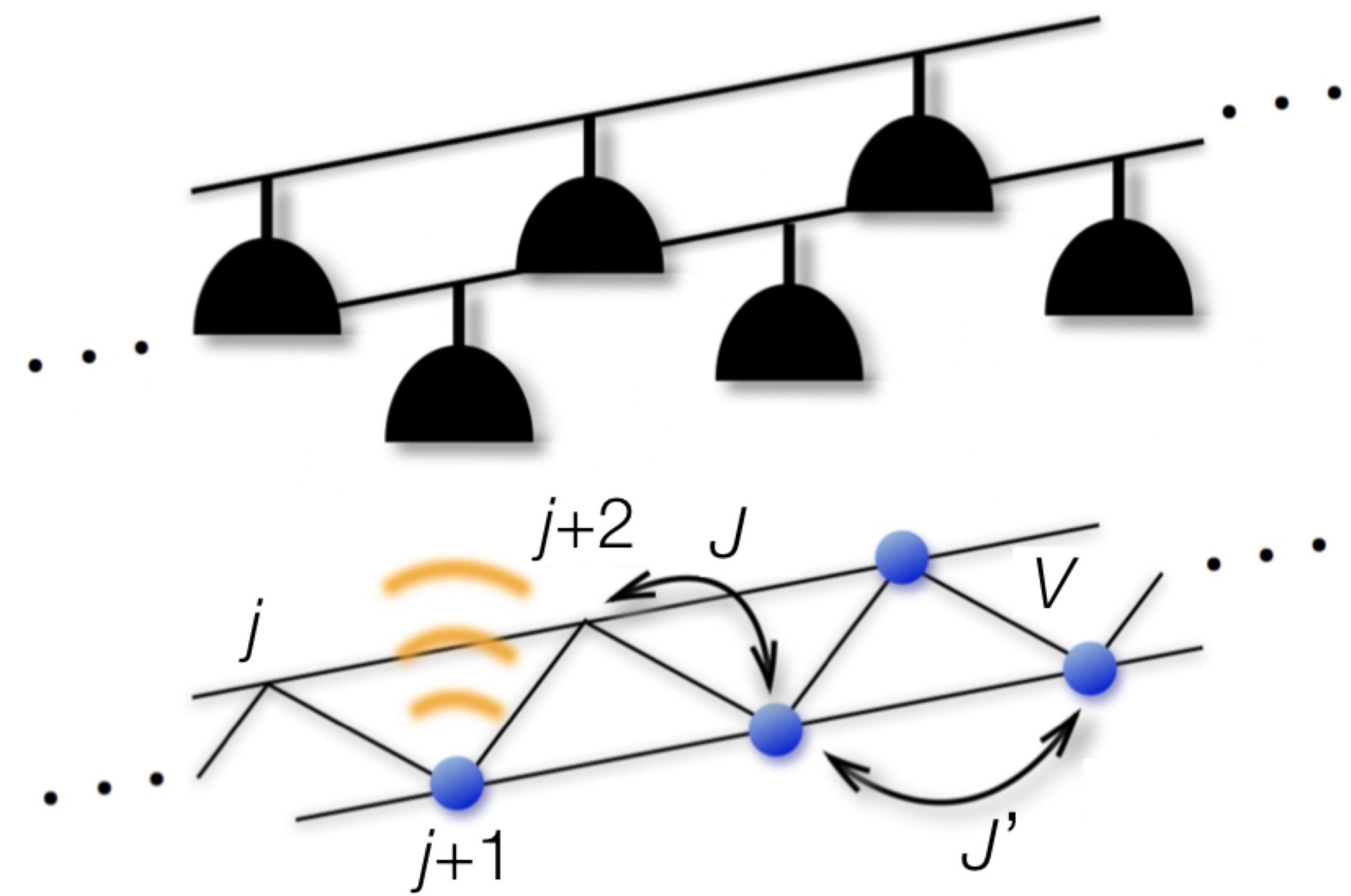}
\caption{A hard-core boson chain subject to continuous position measurement via light scattering. 
The lattice site occupied by a boson is illustrated by a blue filled circle and $j$ labels the lattice site. 
The (next-)nearest-neighbor hopping is represented by ($J'$) $J$ while the  interaction strength is $V$.}
\label{fig:schem}
\end{figure}

We consider a hard-core boson chain, or equivalently a spin-1/2 chain, of the length $L$ given by the Hamiltonian,
\begin{align} \label{eq:XXZHam}
H &= \sum_{j=1}^L \biggl[ \frac{J}{2} \left( b^\dagger_j b_{j+1} +b_j b^\dagger_{j+1} \right) +V n_j n_{j+1} \nonumber \\ 
&+\frac{J'}{2} \left( b^\dagger_j b_{j+2} +b_j b^\dagger_{j+2} \right) \biggr], 
\end{align}
where $b_j$ ($b^\dagger_j$) is the bosonic annihilation (creation) operator acting on site $j$, which is subject to the hard-core constraint $(b_j)^2=0$, and $n_j = b^\dagger_j b_j$. 
The model is schematically illustrated in Fig.~\ref{fig:schem}. 
For the case with the nearest-neighbor hopping only (i.e., $J'=0$), the Hamiltonian becomes the well-known, integrable XXZ chain, which can be further mapped onto a free fermion chain for $V=0$ via the Jordan-Wigner transformation. 
A nonzero next-nearest-neighbor hopping $J'$ breaks the integrability. 
Throughout this paper, we adopt the periodic boundary condition: $b_{L+j} \equiv b_j$. 

\subsection{Conditional dynamics under continuous measurement}
\label{sec:QTDynamics}

We consider the nonunitary dynamics of the model \eqref{eq:XXZHam} under measurement on the basis of the quantum trajectory approach \cite{Daley14}, whose applications to many-body quantum systems can be found in the literature \cite{Pichler10, Schachenmayer14, Mazzucchi16, Wade16, Sorensen18, Ashida18}.
To proceed, we start with the Lindblad master equation, which describes unconditional, dissipative dynamics of the model coupled to a Markovian environment and is given by
\begin{align} \label{eq:MasterEq}
\frac{d}{dt} \rho (t) &= -i [H, \rho(t)] \nonumber \\
&+\sum_j \gamma_j \left( L_j \rho(t) L^\dagger_j -\frac{1}{2} L^\dagger_j L_j \rho(t) -\frac{1}{2} \rho(t) L^\dagger_j L_j \right),
\end{align}
where $\rho(t)$ is the density matrix. 
The continuous measurement of local occupation numbers via, e.g., dispersive light scattering can be modeled by choosing the Lindblad operators to be
\begin{align}
L_j = n_j.
\end{align}
This relation has been derived from the microscopic light-matter interactions for experimentally relevant setups of ultracold atoms in optical lattice both at the levels of master equation \cite{Pichler10} and continuous monitoring \cite{Ashida15}.

Writing the basis states as $n_j | N \rangle_j = N | N \rangle_j$ ($N=0,1$), the number operator $n_j$ is obviously a projection operator onto an occupied state $| 1 \rangle$ at site $j$, and thus $n^2_j = n_j$. 
The constant $\gamma_j$ controls the strength of the measurement at site $j$, and $1/\gamma_j$ is a characteristic time scale called the Zeno time \cite{BP}. 
Heating dynamics and the subsequent thermalization of the related models have previously been studied in Refs.~\cite{Schachenmayer14, Yanay14, Ashida18, Bernier20}.
For our purpose of examining the entanglement entropy, we need to go beyond the unconditional dynamics of the model for which the mixed-state density matrix averaged over measurement outcomes is analyzed. 
Instead, we focus on a quantum trajectory $| \psi (t) \rangle$, or said differently, a pure-state density matrix $\rho(t) = | \psi(t) \rangle \langle \psi(t) |$, which describes the dynamics conditioned on measurement outcomes without averaging \cite{Dalibard92, Ueda92, Carmichael}.

The trajectory dynamics for a free-fermion chain ($V=J'=0$) under continuous position measurement has been studied in Ref.~\cite{Cao19} where each realization of quantum trajectory $| \psi(t) \rangle$ is generated by solving the stochastic Schr\"{o}dinger equation obeying the Wiener process \cite{BP}. 
In this study, we instead solve the stochastic Schr\"{o}dinger equation obeying the discrete stochastic process, known as the marked point process, which is expressed as \cite{BP}
\begin{align} \label{eq:SSE}
d | \psi(t) \rangle &= -i \left( H_\textrm{eff} +\frac{i}{2} \sum_{j=1}^L \gamma_j \| L_j | \psi(t) \rangle \|^2 \right) | \psi(t) \rangle dt \nonumber \\
&+\sum_{j=1}^L \left( \frac{L_j | \psi(t) \rangle}{\| L_j | \psi(t) \rangle \|} -| \psi(t) \rangle \right) dN_j(t). 
\end{align}
Here, $H_\textrm{eff}$ is the non-Hermitian Hamiltonian defined by
\begin{align} \label{eq:NonHermitianHam}
H_\textrm{eff} = H -\frac{i}{2} \sum_{j=1}^L \gamma_j L_j L^\dagger_j, 
\end{align}
and $dN_j(t)$ are discrete random variables that take mean values, 
\begin{align}
E[dN_j(t)] &= \gamma_j \| L_j | \psi(t) \rangle \|^2 dt,
\end{align}
with $E[\cdots]$ representing the ensemble average over the stochastic process, and that satisfy the stochastic calculus: 
\begin{align}
dN_j(t) dN_k(t) = \delta_{jk} dN_j(t).
\end{align}
Specifically, the stochastic Schr\"{o}dinger equation can in practice be simulated by the following update algorithm \cite{BP}:

\begin{enumerate}
\item 
Starting from an initial state $| \psi(0) \rangle$ at $t=0$, we evolve the state by the Schr\"{o}dinger equation with the non-Hermitian Hamiltonian $H_\textrm{eff}$, 
\begin{align} \label{eq:NonHermitianSE}
\frac{d}{dt} | \psi(t) \rangle = -iH_\textrm{eff} | \psi(t) \rangle, 
\end{align}
until a waiting time $t=\tau$ set by the relation, 
\begin{align} \label{eq:WaitingTime}
\| e^{-iH_\textrm{eff} \tau} | \psi(t) \rangle \|^2 = \eta, 
\end{align}
where $\eta$ is chosen from the uniform distribution in $[0,1]$. 

\item At $t=\tau$, a quantum \emph{jump} by the Lindblad operator $L_j$ occurs with the probability $p_j$ given by 
\begin{align} \label{eq:JumpProbability}
p_j = \frac{\gamma_j \| L_j | \psi(\tau) \rangle \|^2}{\sum_k \gamma_k \| L_k | \psi(\tau) \rangle \|^2}.
\end{align}
The state $\left| \psi(\tau) \right>$ is then replaced as
\begin{align} \label{eq:Jump}
| \psi(\tau) \rangle \to \frac{L_j | \psi(\tau) \rangle}{\| L_j | \psi(\tau) \rangle \|}.
\end{align}
\item Process 1 is repeated by replacing the initial state $| \psi(0) \rangle$ by Eq.~\eqref{eq:Jump}.
\end{enumerate}
As seen from Eq.~\eqref{eq:WaitingTime}, the waiting-time distribution for quantum jumps or projective measurements is determined by the norm of a state evolved under the nonunitary time evolution and is not ensured to be uniform in general.

In the following discussion, we consider the continuous measurement performed uniformly in space, that is, 
\begin{align}
\gamma_j \equiv \gamma.
\end{align} 
In this case, the nonunitary time evolution by the non-Hermitian Hamiltonian \eqref{eq:NonHermitianHam} becomes equivalent to the unitary dynamics by the (Hermitian) Hamiltonian $H$ upon the normalization of the state norm:
\begin{align}
H_\textrm{eff} &= H -\frac{i\gamma}{2} \sum_{j=1}^L n_j \\
&= H -\frac{iL \gamma \nu}{2}. 
\end{align}
Here, we have replaced the non-Hermitian part by filling $\nu = \langle n_\textrm{tot} \rangle/L$ of the initial state in the second line, since the total particle number $n_\textrm{tot} = \sum_{j=1}^L n_j$ is conserved under the time-evolution by $H_\textrm{eff}$ and the projective measurements of $n_j$. 
Therefore, combined with Eq.~\eqref{eq:WaitingTime}, the projective measurement is performed at a constant rate:
\begin{align}
\tau = -\frac{\ln \eta}{L \gamma \nu}.
\end{align}

In this marked-point-process formulation of quantum trajectory, the corresponding dynamics might be compared with the unitary circuit dynamics with random projective measurements, apart from the fact that the unitary dynamics is given by the unitary time evolution with the Hamiltonian \eqref{eq:XXZHam} in the former case. 
In the unitary circuit dynamics, the probability of measurement follows a uniform distribution in both space and time. 
For our case, the probability distribution of measurement is uniform in time but not in space as given in Eq.~\eqref{eq:JumpProbability}. 
Since $\| n_j |\psi(\tau) \rangle \|^2 = \langle \psi (t) | n_j | \psi(t) \rangle$, the probability distribution in space depends on the expectation values of occupation numbers $n_j$ at time $t=\tau$ when a quantum jump is to be observed. 
This indicates that occupied states are more likely to be measured than empty states, measured states too at the next quantum jump, and thus the particle configuration of a quantum trajectory might be kept frozen in that of the initial state during the time evolution. 
However, as we will see in the next section, the expectation values of the occupation numbers $n_j$ relax to the mean value $\nu$ in a shorter time scale than the waiting time $\tau$ for quantum jumps in the regime we are interested in. 
Hence, a memory of the inhomogeneous particle configuration in the initial state is washed out and the particle density distribution becomes uniform \emph{on average} after a certain time.
This makes the probability distribution of measurement in our model effectively uniform in space, and therefore we expect measurement-induced transitions similar to unitary circuit models for steady-state properties in the quantum trajectory dynamics.

We remark that, in the unconditional dynamics governed by the Lindblad master equation \eqref{eq:MasterEq}, a mixed-state density matrix simply converges to the infinite-temperature state in the long-time limit. In fact, when a jump operator is Hermitian (or more generally normal), the steady state of a generic nonintegrable many-body system should be the infinite-temperature state \cite{Pichler10, Schachenmayer14, Ashida18, RodriguezChiacchio18}. This fact can be inferred from  the facts that the master equation~\eqref{eq:MasterEq} permits the solution $\rho \propto 1$ when $L_j$ is Hermitian (or normal) for all $j$ and that this is the unique solution because of the Perron-Frobenius theorem \cite{SJ76}, unless the ergodicity of the Liouvillean matrix is violated due to additional symmetries \cite{Tindall19}. 

\section{Results}
\label{sec:Results}

In the following, we show numerical results obtained for the Hamiltonian \eqref{eq:XXZHam} with the system sizes from $L=8$ to $24$ under the periodic boundary condition. 
We choose the initial state in the ``N\'{e}el'' state with fixed filling $\nu=1/2$: 
\begin{align} \label{eq:NeelState}
| \psi (0) \rangle = | 0 1 0 1 \cdots 0 1 \rangle.
\end{align}
We set $J=1$ to be the unit of energy. 
For the nonunitary time evolution given in Eq.~\eqref{eq:NonHermitianSE}, which is now equivalent to the unitary time evolution by the Hamiltonian \eqref{eq:XXZHam} (upon the renormalization) as we discussed, we employ the standard RK4 method to solve the Schr\"{o}dinger equation with the maximal time slice $\Delta t =0.01$. 
The time evolutions are performed in the range of $t \in [0,100]$ for nonintegrable cases of (i) $V=0$, $J'=0.5$, (ii) $V=0$, $J'=1$, and (iii) $V=1$, $J'=0.5$, whereas $t \in [0,200]$ for an integrable case of (iv) $V=1$, $J'=0$. 
The averaged quantities are computed with 400--800 realizations of the quantum trajectories. 

\subsection{Particle density profiles}
\label{sec:ParticleDensity}

\begin{figure}
\includegraphics[clip,width=0.48\textwidth]{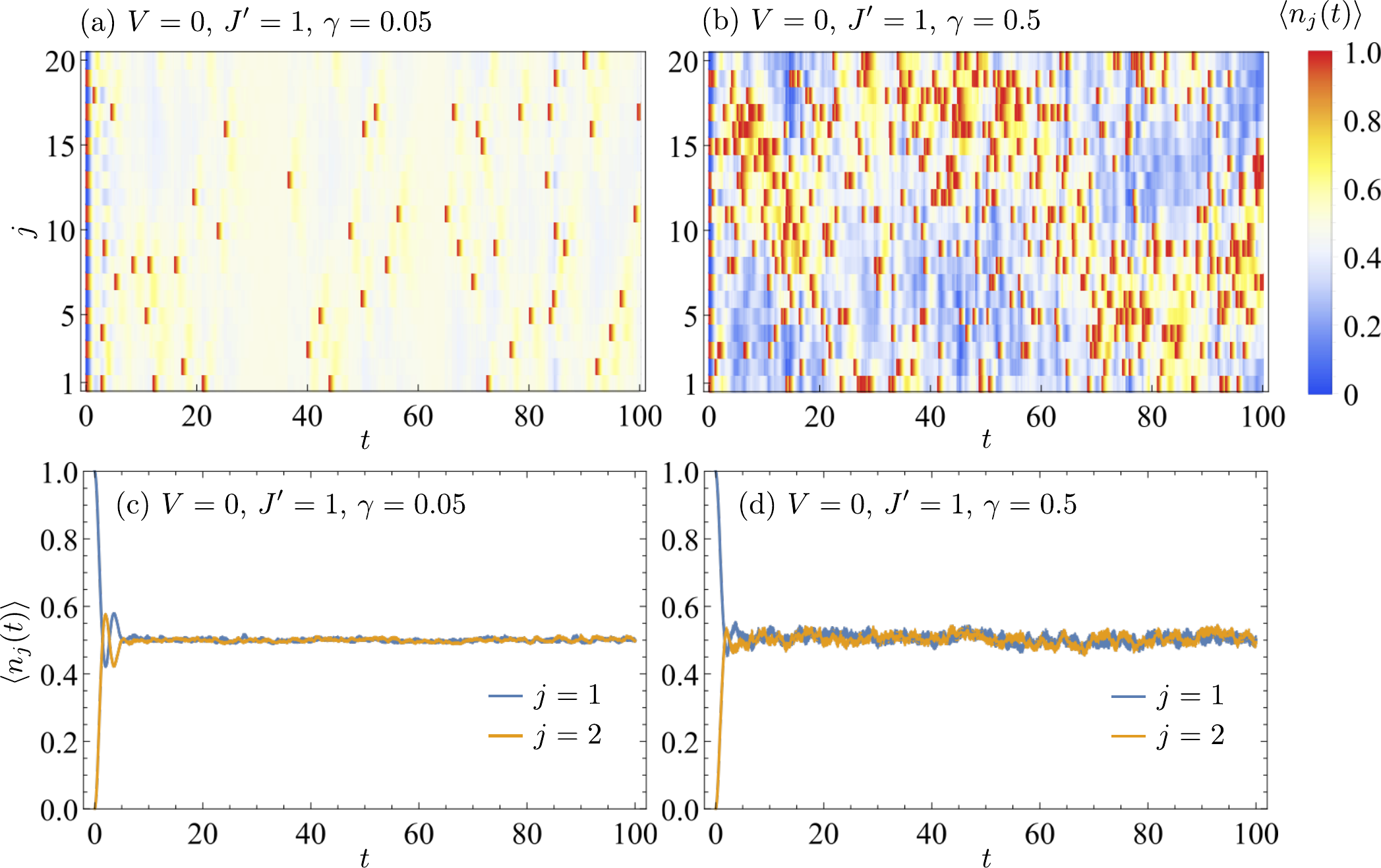}
\caption{Time evolutions of the local particle densities $\langle n_j(t) \rangle$ for a nonintegrable model with $V=0$, $J'=1$, and $L=20$ at half filling. 
The panels (a) and (b) show those for a \emph{single} trajectory with $\gamma=0.05$ and $\gamma=0.5$, respectively. 
The panels (c) and (d) show their \emph{averages} over 400 quantum trajectories with standard error for sites $j=1$ and $j=2$.}
\label{fig:NdisTdep1}
\end{figure}

\begin{figure}
\includegraphics[clip,width=0.48\textwidth]{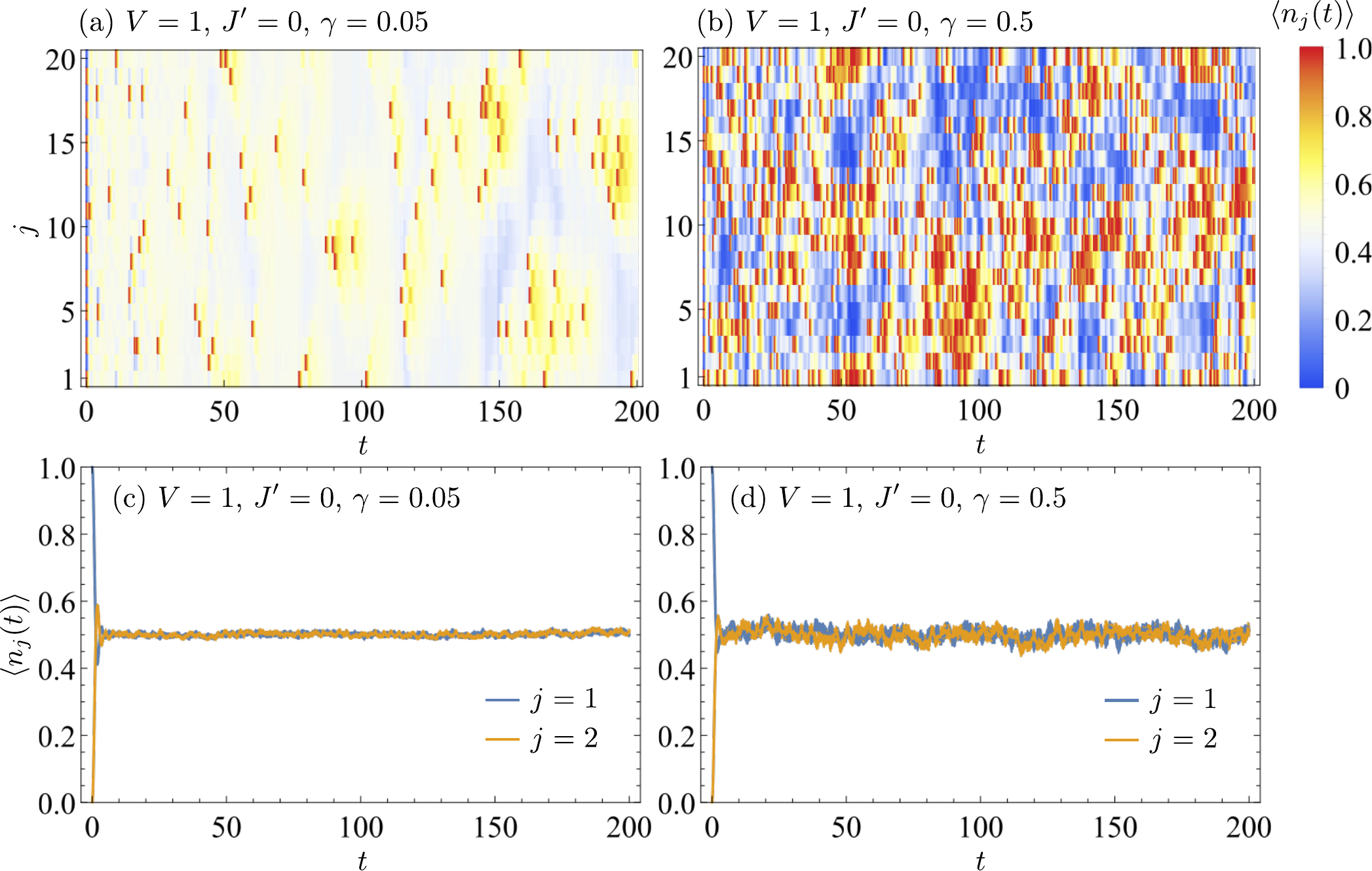}
\caption{Time evolutions of the local particle densities $\langle n_j(t) \rangle$ for an integrable model with $V=1$, $J'=0$, and $L=20$, as shown in Fig.~\ref{fig:NdisTdep1}.}
\label{fig:NdisTdep2}
\end{figure}

\begin{figure}
\includegraphics[clip,width=0.48\textwidth]{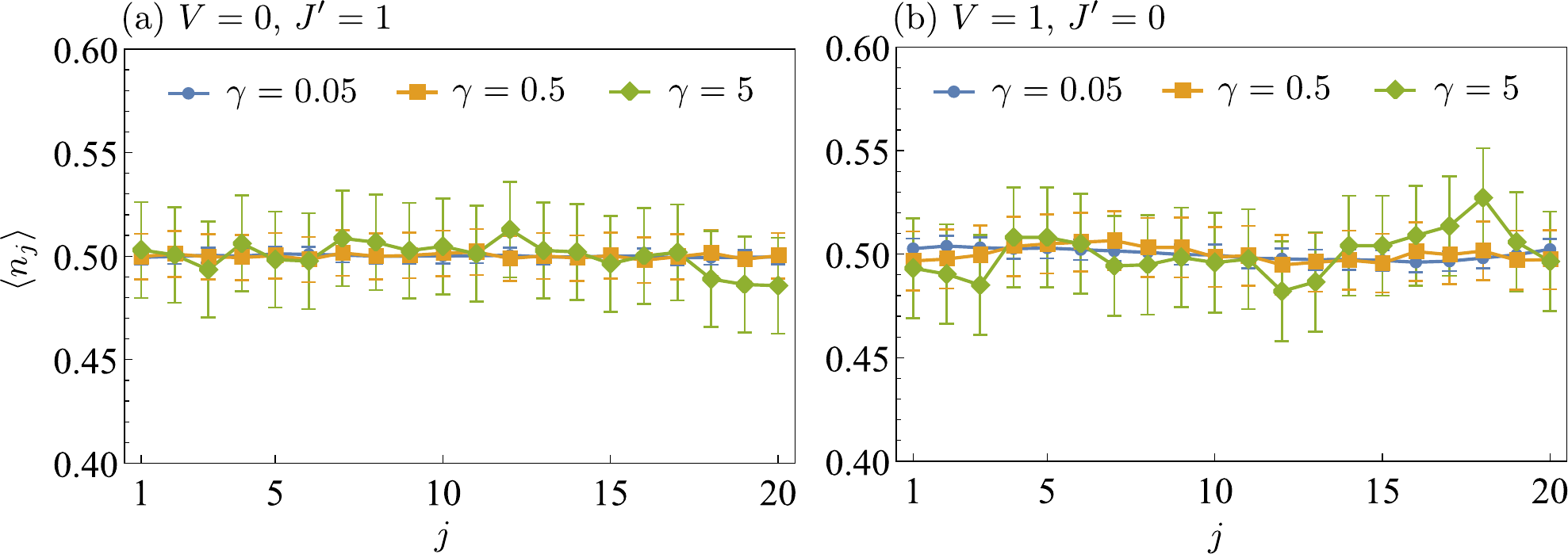}
\caption{Steady-state values of the particle densities $\langle n_j \rangle$ for the $L=20$ systems of (a) a nonintegrable case with $V=0$, $J'=1$ and (b) an integrable case with $V=1$, $J'=0$ for $\gamma=0.05,0.5,5$. 
The averages are taken over 400 quantum trajectories and over the time interval (a) $t \in [50,100]$ or (b) $t \in [150,200]$.}
\label{fig:NsqrtdisSS}
\end{figure}

We first show the spatiotemporal evolutions of the local particle densities for a nonintegrable model ($V=0$, $J'=1$) for the measurement strengths $\gamma=0.05$ and $\gamma=0.5$ in Fig.~\ref{fig:NdisTdep1}. 
For a small $\gamma$, the particle density $\langle n_j(t) \rangle = \langle \psi(t) | n_j | \psi(t) \rangle$ for a \emph{single} quantum trajectory $| \psi(t) \rangle$, which is initially fixed to 0 or 1, relaxes to the filling $\nu=1/2$ in time $t \sim O(1)$ only with small deviations [Fig.~\ref{fig:NdisTdep1} (a)]. 
Although it jumps to $1$ right after the projective measurement occurs, it then relaxes to $1/2$ again by the (non)unitary time evolution.
Such relaxation dynamics has been observed previously in a similar setup \cite{Ashida18}, but we are now interested in a regime where the relaxation and measurement rates become more competitive.
For a large $\gamma$, the density profiles become more diffusive due to competition between the (non)unitary time evolutions and the projective measurements [Fig.~\ref{fig:NdisTdep1} (b)]. 
However, the particle densities \emph{averaged} over quantum trajectories clearly converge to $\nu=1/2$ for a late time, irrespective of their initial values, as seen from Figs.~\ref{fig:NdisTdep1} (c)-(d). 
We also find similar spatiotemporal behaviors of the particle densities for an integrable model ($V=1$, $J'=0$) as shown in Fig.~\ref{fig:NdisTdep2}, although their relaxation dynamics is commonly recognized to be more subtle than that for nonintegrable models. 
In Fig.~\ref{fig:NsqrtdisSS}, we show the \emph{steady-state} values of the particle densities, $\langle n_j \rangle$, which are obtained by averaging $\langle n_j(t) \rangle$ over both quantum trajectories and the time interval $t \in [50,100]$ for the nonintegrable case ($V=0$, $J'=1$) and over $t \in [150,200]$ for the integrable case ($V=1$, $J'=0$), in which $\langle n_j(t) \rangle$ appears to saturate.
The steady-state values of $\langle n_j \rangle$ take the value of filling $\nu=1/2$ within error, irrespective of the particle position $j$, and thus keep no memories of the initial particle configuration of the N\'{e}el state. 
This is expected from the fact that the steady-state particle density $\langle n_j(t \to \infty) \rangle$ averaged over quantum trajectories should coincide with that of the infinite-temperature mixed state obtained from the master equation. 
This indicates that the distribution of the particle densities and thereby the probability distribution $p_j$ of the projective measurements are uniform in space, on average, for a late time. 
In this sense, our system is considered to be not very far from the unitary dynamics with random projective measurements, in which the probability distribution of the measurement is uniform in both space and time. 
This also ensures the spatial translational invariance of steady-state quantities averaged over quantum trajectories in the following analysis.

\subsection{Entanglement profiles}

\begin{figure}
\includegraphics[clip,width=0.48\textwidth]{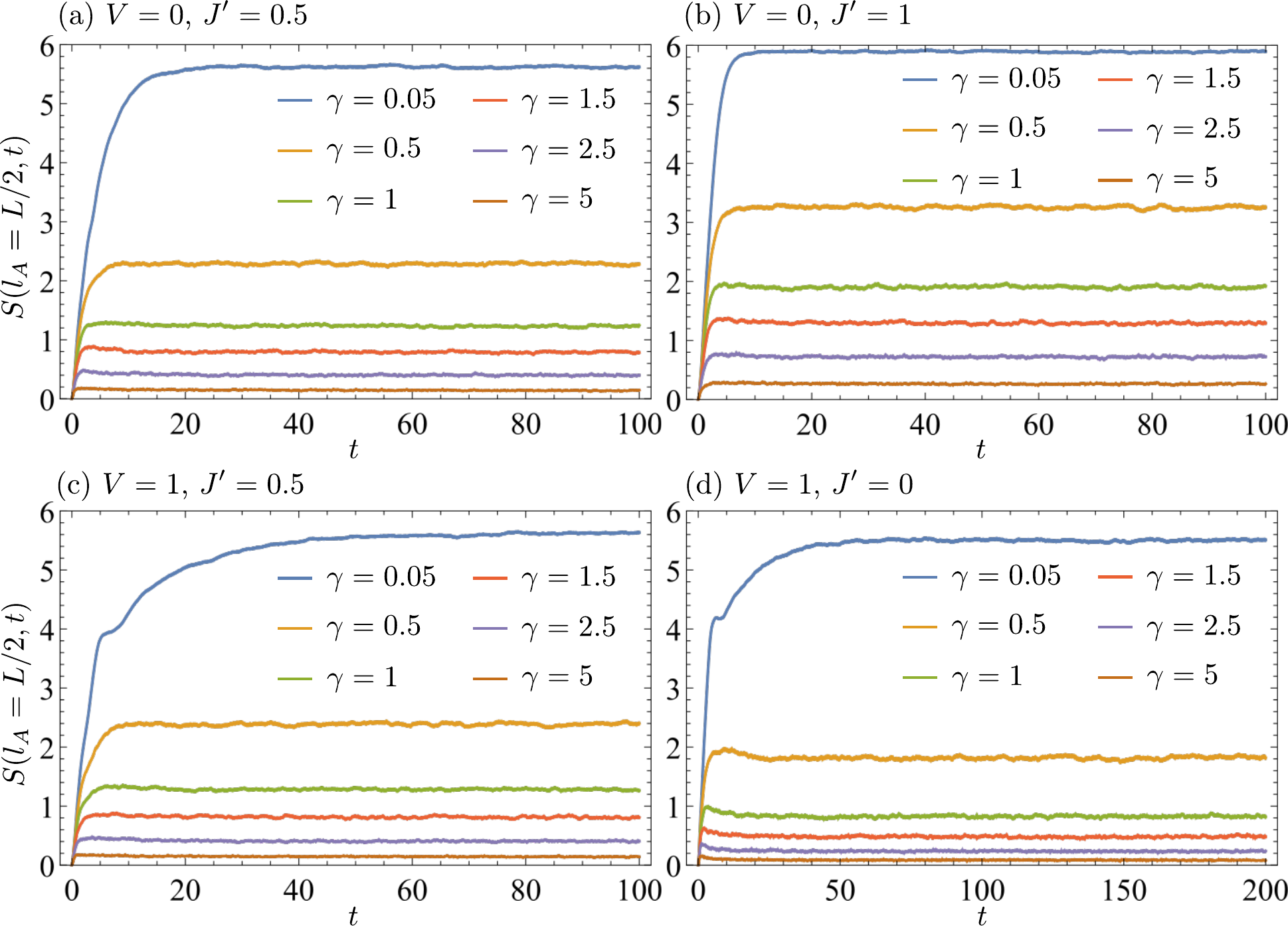}
\caption{Time evolutions of the von Neumann entanglement entropies $S(l_A=L/2,t)$ averaged over 800 quantum trajectories for several values of the measurement strength $\gamma$. 
The data are shown for the $L=20$ systems of nonintegrable models (a) $V=0$, $J'=0.5$, (b) $V=0$, $J'=1$, and (c) $V=1$, $J'=0.5$, and an integrable model (d) $V=1$, $J'=0$. 
The standard errors are smaller than the widths of the curves.}
\label{fig:EntTdep}
\end{figure}

\begin{figure}
\includegraphics[clip,width=0.35\textwidth]{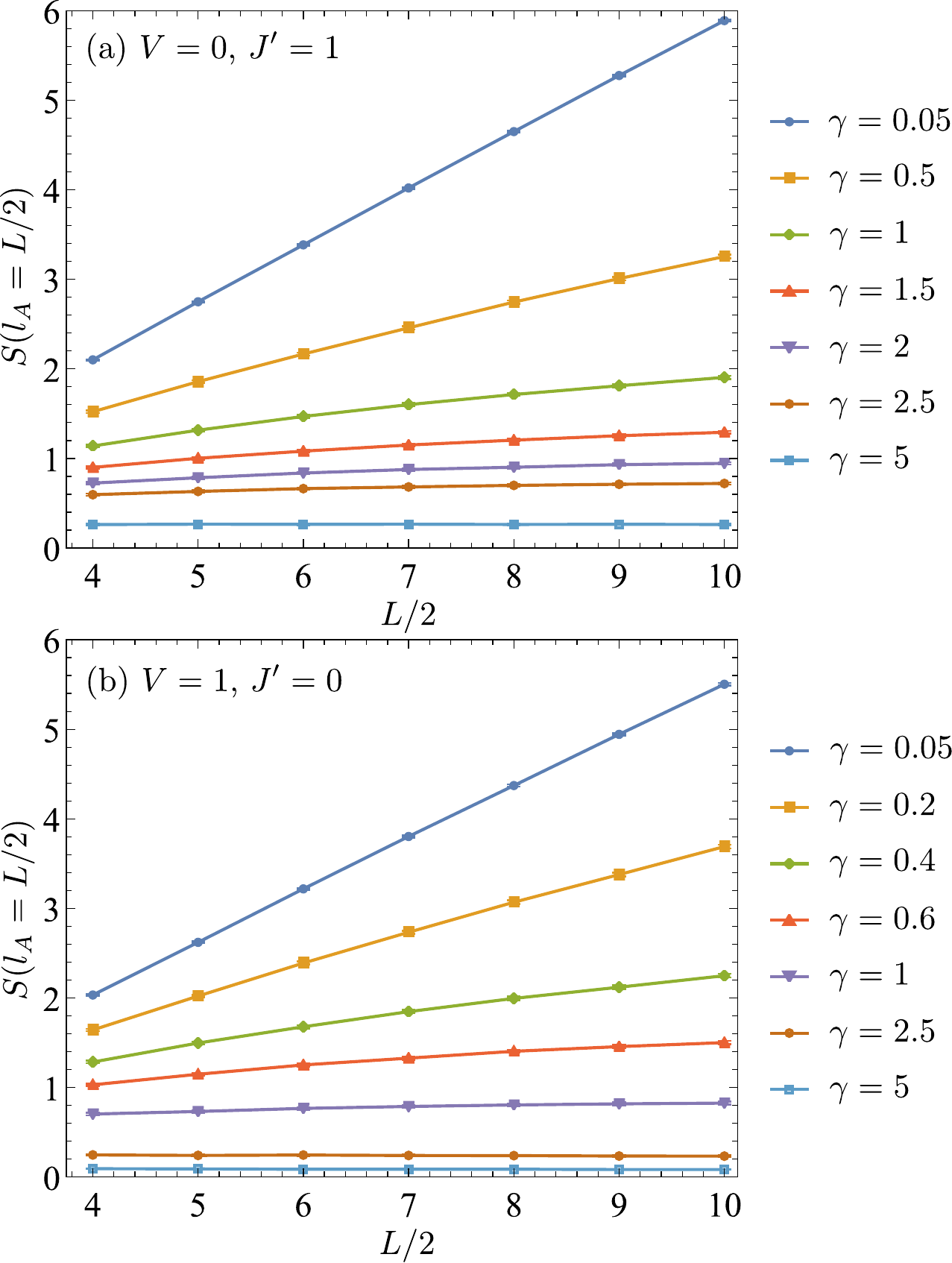}
\caption{Steady-state values of the von Neumann entanglement entropies $S(l_A=L/2)$ plotted against half of the system size $L$ for several values of the measurement strength $\gamma$. 
The averages are taken over 800 quantum trajectories and over the time intervals (a) $t \in [50,100]$ for a nonintegrable model with $V=0$, $J'=1$ and (b) $t \in [150,200]$ for an integrable model with $V=1$, $J'=0$.}
\label{fig:EntSSLdep}
\end{figure}

We now move on to our main focus on the quantum trajectory dynamics of the pure-state entanglement under continuous measurement. 
We quantify the entanglement under bipartition of the system by the von Neumann entanglement entropy, 
\begin{align}
S_A(t) = -\textrm{Tr}_A [\rho_A(t) \ln \rho_A(t)], 
\end{align}
where $\rho_A(t)$ is the reduced density matrix of the subregion $A$ obtained by tracing out the degrees of freedom in the complement of $A$, denoted by $\bar{A}$, from the density matrix $\rho(t) = | \psi(t) \rangle \langle \psi(t) |$ as
\begin{align}
\rho_A(t) = \textrm{Tr}_{\bar{A}} \, \rho(t).
\end{align}
Owing to the translational invariance of $S_A(t)$ averaged over quantum trajectories, we choose $A=\{ j \, | \, 1 \leq j \leq l_A\}$ with $l_A$ being the length of $A$ without loss of generality and write $S_A(t) = S(l_A,t)$.
In Fig.~\ref{fig:EntTdep}, we show time evolutions of the half-chain entanglement entropies $S(l_A=L/2,t)$ averaged over quantum trajectories.
While the entanglement entropy is initially zero as the states are prepared in the N\'{e}el state as a product state, the averaged ones linearly increase in early time and saturate to steady-state values in late time, which decrease with increasing of the measurement strength $\gamma$. 
We then obtain the steady-state entanglement entropies $S(l_A)$ by averaging $S(l_A,t)$ over the time interval $t \in [50,100]$ for nonintegrable models and $t \in [150,200]$ for integrable models. 
In Fig.~\ref{fig:EntSSLdep}, we show the system-size dependence of the half-chain entanglement entropies $S(l_A=L/2)$ in these steady-state regimes.
For both nonintegrable ($V=0$, $J'=1$) and integrable ($V=1$, $J'=0$) cases, the entanglement entropy appears to increase linearly in the system size for small $\gamma$, indicating the volume-law entanglement expected for thermalizing states \cite{Ashida18}, while its increment is dramatically suppressed for large $\gamma$, signaling the area-law entanglement caused by frequent applications of the projective measurement. 
Thus, the results indicate the signature of an entanglement transition at which the entanglement entropy can scale logarithmically as $\ln L$. 
However, we find that it is in practice difficult to accurately estimate critical measurement strengths $\gamma_c$ by just fitting functional forms of the entanglement entropy with small-system-size data. 
Instead, we adopt the mutual information proposed in Ref.~\cite{Li19} as a more convenient measure to quantify the critical value $\gamma_c$ in the following subsection.
We emphasize that such a steady-state property sensitive to the measurement strength is a unique consequence of the conditional dynamics at the single-trajectory level, far beyond the unconditional dynamics at the level of the master equation \eqref{eq:MasterEq}.
This unique feature is particularly intriguing in light of the fact that the latter dynamics merely ends up with the featureless, infinite-temperature mixed state (see also discussions in  Sec.~\ref{sec:QTDynamics}). 

\subsection{Mutual information and correlation functions}
\label{sec:MutualInf}

For two subregions $A$ and $B$ embedded in the whole system, the von-Neumann mutual information is defined by
\begin{align} \label{eq:MutualInf}
I_{AB}(t) = S_A(t) + S_B(t) -S_{A \cup B}(t), 
\end{align}
where $S_A$, $S_B$, and $S_{A \cup B}$ are the von Neumann entanglement entropies of the subsystems $A$ and $B$ and their disjoint union $A \cup B$, respectively. 
Here, we choose the subregions $A$ and $B$ to be single sites separated by the distance $r_{AB}$ on a ring of the length $L$. 
Because of spatial homogeneity of the late-time dynamics as discussed in Sec.~\ref{sec:ParticleDensity}, we may choose $A= \{ 1 \}$, $B=\{ r_{AB}+1 \}$, and thus $A \cup B = \{ 1, r_{AB}+1 \}$ without loss of generality, and the mutual information can be seen as a function of $r_{AB}$; $I_{AB}(t) = I_{AB}(r_{AB},t)$.
For unitary circuits with random projective measurements, the mutual information $I_{AB}(r_{AB},t)$ in a steady-state regime has been found to draw a peak at the measurement-induced entanglement transition \cite{Li19}. 
We confirm that this is also the case for our quantum trajectory dynamics. 

\begin{figure*}
\includegraphics[clip,width=\textwidth]{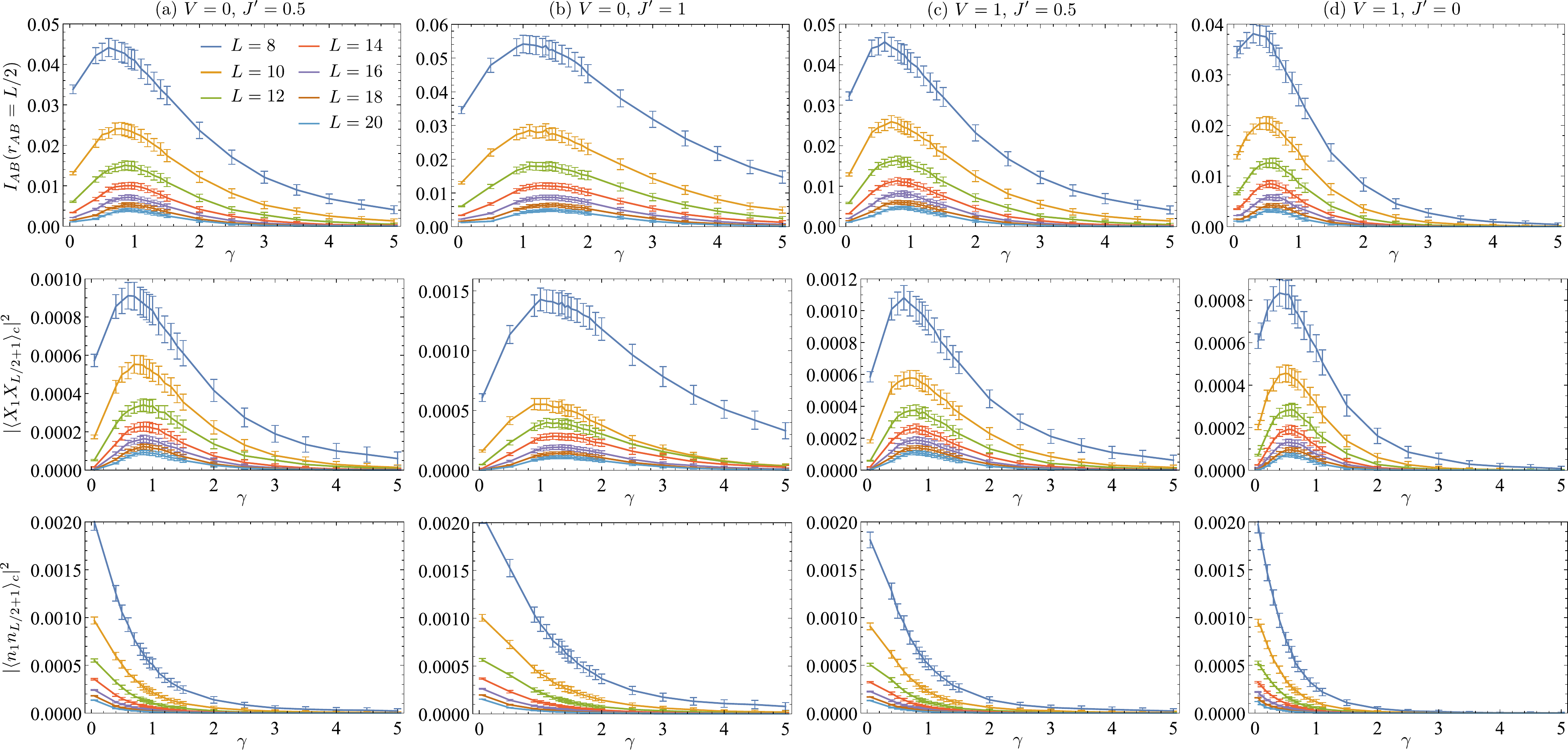}
\caption{Steady-state values of the mutual information $I_{AB}(r_{AB}=L/2)$ and the absolute squares of the connected correlation functions $\langle X_1 X_{L/2+1} \rangle_c$ and $\langle n_1 n_{L/2+1} \rangle_c$ between antipodal sites are plotted as functions of $\gamma$. 
The data are shown for $L=8$ to $20$ and for the parameter choices of nonintegrable models (a) $V=0$, $J'=0.5$, (b) $V=0$, $J'=1$, and (c) $V=1$, $J'=0.5$, and an integrable model (d) $V=1$, $J'=0$. 
The averages are taken over 800 quantum trajectories and over the time intervals $t \in [50,100]$ for (a)--(c) and $t \in [150,200]$ for (d).}
\label{fig:MutualInfSS}
\end{figure*}

As before, we define the steady-state quantities by averaging over the time interval $t \in [50,100]$ for nonintegrable models and $t \in [150,200]$ for integrable models, besides averaging over quantum trajectories.
The top panels of Fig.~\ref{fig:MutualInfSS} show the steady-state values of the mutual information $I_{AB}(r_{AB}=L/2,t)$ between antipodal sites $j=1$ and $j=L/2+1$ on a ring. 
They indeed draw broad peaks as the measurement strength $\gamma$ is varied. 
The peak positions appear to eventually converge with increasing the system size; although these estimates are not quite accurate due to statistical error, the peak positions $\gamma_c$ roughly read off from the $L=20$ systems as $\gamma_c \sim 0.9$ for $(V,J')=(0,0.5)$ and $(1,0.5)$, $\gamma_c \sim 1.4$ for $(V,J')=(0,1)$, and $\gamma_c \sim 0.6$ for $(V,J')=(1,0)$. 
The validity of these estimates is substantiated by the scaling analysis in the next subsection.

We have also computed the steady-state values of the connected correlation functions of operators defined by
\begin{align}
X_j \equiv \frac{1}{2}(b_j+b_j^\dagger)
\end{align}
and of the particle number operators $n_j$ between antipodal sites, whose absolute squares are shown in the two bottom panels of Fig.~\ref{fig:MutualInfSS}.
The correlation functions are closely related to the mutual information in the following sense. 
It is known that the von Neumann mutual information \eqref{eq:MutualInf} gives an upper bound on correlation functions \cite{Wolf08}, 
\begin{align}
I_{AB} \geq \frac{| \langle O_A O_B \rangle_c|^2}{2\| O_A \|^2 \| O_B \|^2},
\end{align}
where $O_A$ and $O_B$ are arbitrary operators supported on the subregions $A$ and $B$, respectively, $\langle O_A O_B \rangle_c \equiv \langle O_A O_B \rangle -\langle O_A \rangle \langle O_B \rangle$ is the connected correlation function, and $\| \cdots \|$ denotes the Euclidean operator norm equivalent to the largest singular value of the operator to be evaluated. 
It has been found in Ref.~\cite{Li19} that the absolute square of the correlation function of the Pauli operators, $|\langle Z_1 Z_{r_{AB}+1} \rangle_c|^2$, shows a peak structure similar to the mutual information for unitary circuits with random projective measurements, in which case $Z_j$ itself is related to the measured operator $P_j = (Z_j +1)/2$. 
Compared with our case, we can identify the particle number operator $n_j = P_j$ as $(Z_j+1)/2$ and the operator $X_j$ as another Pauli operator up to normalization of a factor of 2.
We find that the correlation functions of measured operators $n_j$ monotonically decay with increasing of $\gamma$, whereas those of $X_j$ feature peak structures quite similar to the mutual informations. 
As we will see in the next subsection, the correlation functions of $X_j$ even show almost the same exponents as those for the mutual information in their algebraic decays at the entanglement transition. 
Such a distinct behavior in the correlation functions might be attributed to the particle number conservation both in the Hamiltonian and in the entire quantum trajectory dynamics in our model; in the absence of projective measurements, the unitary time evolutions by a nonintegrable Hamiltonian and by random unitary circuits both provide chaotic dynamics and are expected to lead to thermalization, while distinct behaviors could arise once the dynamics preserves some symmetry, e.g., the $U(1)$ symmetry \cite{Khemani18, Rakovszky18}. 
We speculate that the $U(1)$ symmetry associated with the particle number conservation is the origin of the different behavior in the correlation function of the measured operators $P_j=n_j$ from the case of random quantum circuits studied in Ref.~\cite{Li19}. 
It will be an interesting future problem to examine whether such distinct behaviors dependent on the presence or absence of $U(1)$ symmetry can be found in different setups of unitary dynamics with projective measurements.

\subsection{Scaling behaviors at entanglement transitions}
\label{sec:Scaling}

\begin{figure}
\includegraphics[clip,width=0.48\textwidth]{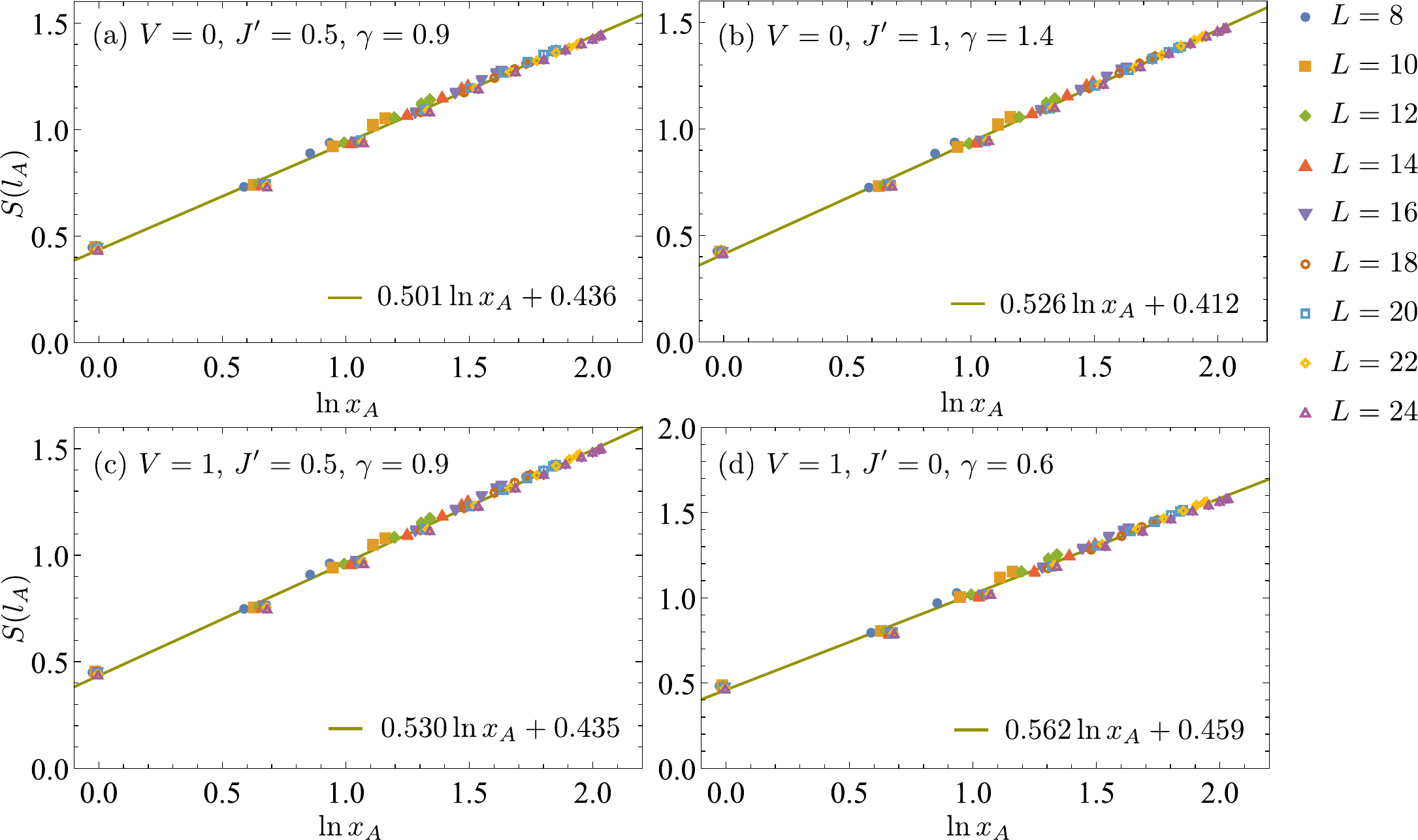}
\caption{Steady-state values of the von Neumann entanglement entropy $S(l_A)$ at the entanglement transitions $\gamma=\gamma_c$ are plotted against the logarithm of the chord length $x_A$ of the subsystem $A$ for $L=8$ to $24$. 
The averages are taken over 400 quantum trajectories and over the time intervals $t \in [50,100]$ for nonintegbrable models (a)--(c) and $t \in [150,200]$ for an integrable model (d). 
The solid lines are fitting functions of the form $\alpha_S \ln x_A +\beta_S$.}
\label{fig:EEScaling}
\end{figure}

We here confirm emergent conformal invariance at the entanglement transitions, which are located by the peak positions of the mutual informations for the $L=20$ systems as discussed above. 
From Ref.~\cite{Calabrese04}, the von Neumann entanglement entropy of a conventional 1D critical system described by CFT under the periodic boundary condition is given by 
\begin{align}
S(l_A) = \frac{c}{3} \ln x_A +c', 
\end{align}
where $x_A = (L/\pi) \sin (\pi l_A/L)$ is the chord length of the subsystem $A$, $c$ is the central charge, and $c'$ is a nonuniversal constant. 
Thus, one may assume the scaling form $S(l_A) = \alpha_S \ln x_A +\beta_S$ right at the entanglement transitions \cite{Skinner19, Li19}. 
We show the steady-state values of the entanglement entropies for different subsystem sizes in Fig.~\ref{fig:EEScaling}, which are fitted well into this scaling form.
The coefficients $\alpha_S$ take similar values ranging from 0.50 to 0.56 for both nonintegrable models [Figs.~\ref{fig:EEScaling} (a)--(c)] and an integrable model [Fig.~\ref{fig:EEScaling} (d)]. 
However, none of them are likely to be compared with the unitary CFT for 1D critical systems as $c>1$ often requires larger Lie group symmetries than $U(1)$. 
Nevertheless, the $U(1)$-symmetry-resolved entanglement exhibits some feature like a free boson CFT or a Tomonaga-Luttinger liquid theory as we discuss in the next subsection.
This may indicate possible unconventional quantum critical phenomena that are distinct from the known universality classes. 
The coefficients $\alpha_S$ are also different from the previously reported values for random unitary circuits \cite{Skinner19, Li19, Zabalo20} and for a Bose-Hubbard model with random projective measurements \cite{Tang20}, which are summarized in Table~\ref{table:Exponents}. 
\begin{table*}
\begin{ruledtabular}
\begin{tabular}{lcccccccc}
Model & $(V,J')=(0,0.5)$ & $(V,J')=(0,1)$ & $(V,J')=(1,0.5)$ & $(V,J')=(1,0)$ & Clifford \cite{Li19} & Haar \cite{Skinner19} & Haar \cite{Zabalo20} & BH \cite{Tang20} \\
\hline
$\alpha_S$ & 0.501(4) & 0.526(3) & 0.530(4) & 0.562(5) & 1.6 & & 1.7(2) & 0.21(1) \\
$\nu$ & 1.34(3) & 1.22(3) & 1.26(3) & 1.21(2) & 1.3 & 2.01(10) & 1.2(2) & 2.00(15) \\
$\Delta$ & 1.270(9) & 1.210(13) & 1.185(9) & 1.273(28) & 2.1 & $\sim 2$ & & 1.29-1.56 \\ 
$\Delta_X$ & 1.290(14) & 1.317(26) & 1.220(7) & 1.257(15) \\
$\Delta_n$ & 1.831(22) & 1.749(28) & 1.706(23) & 1.816(25)
\end{tabular}
\end{ruledtabular}
\caption{Coefficient $\alpha_S$ of the logarithmic part of the entanglement entropy and exponents $\nu$, $\Delta$, $\Delta_X$, and $\Delta_n$ for the entanglement entropy, mutual information, and absolute squares of the correlation functions $\langle X_i X_j \rangle_c$ and $\langle n_i n_j \rangle_c$, respectively, which are obtained for several $V$ and $J'$. 
The results are compared with those for the Clifford \cite{Li19} and Haar \cite{Skinner19, Zabalo20} random unitary circuits and the Bose-Hubbard (BH) model \cite{Tang20} with random projective measurements.}
\label{table:Exponents}
\end{table*}

\begin{figure}
\includegraphics[clip,width=0.48\textwidth]{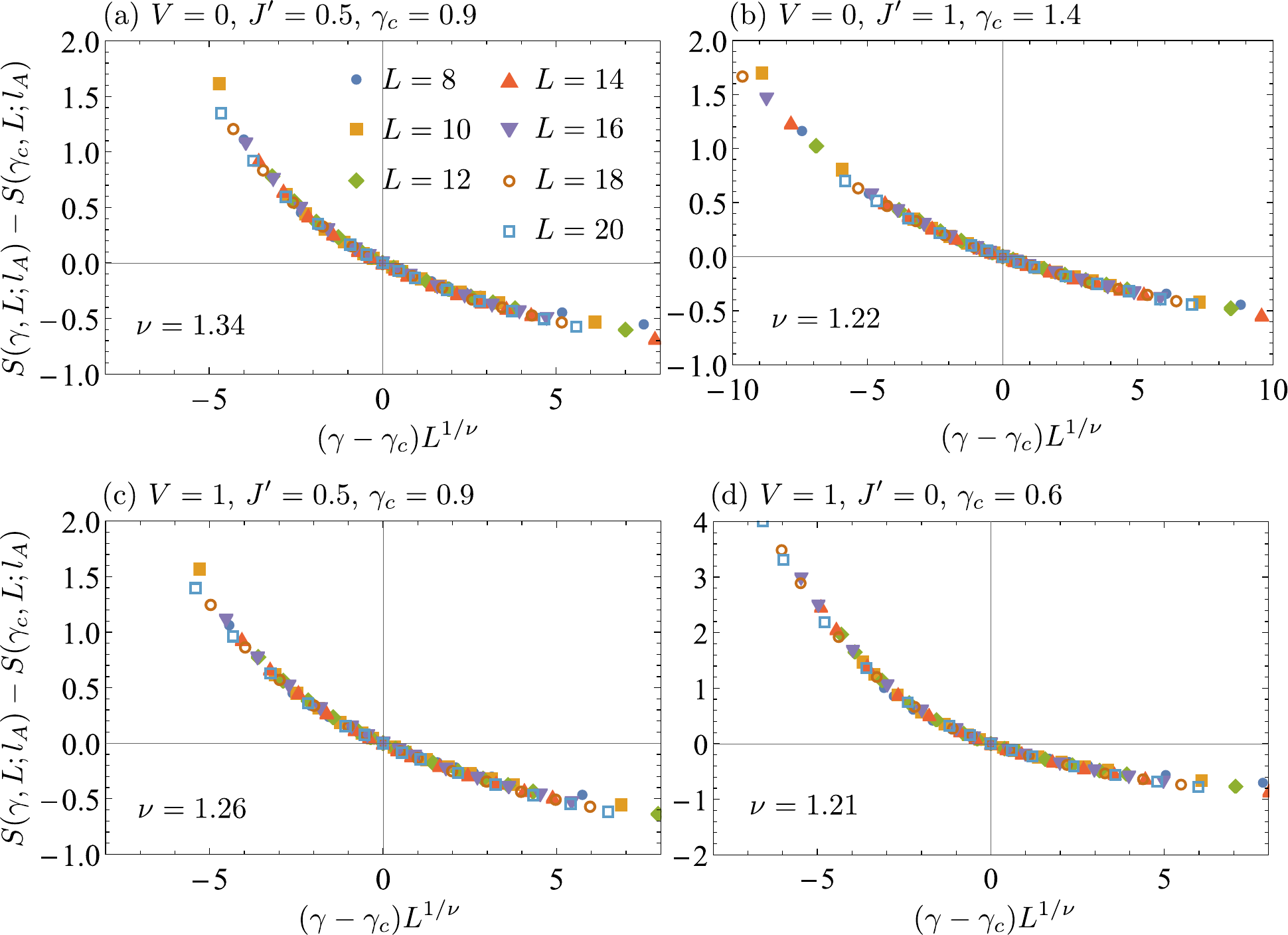}
\caption{Data collapses of the steady-state entanglement entropy into the scaling form \eqref{eq:ScalingEE} for given estimates of the critical measurement strength $\gamma_c$. 
The data are shown for nonintegrable models (a)--(c) and an integrable model (d).
Each data point is the averaged value over 800 quantum trajectories.}
\label{fig:ScalingCollapse}
\end{figure}

In the vicinity of the transition $\gamma \sim \gamma_c$, the entanglement entropy has been proposed to follow the scaling form \cite{Skinner19,Li19}, 
\begin{align} \label{eq:ScalingEE}
S(\gamma,L;l_A=aL)-S(\gamma_c,L;l_A=aL) = F[(\gamma-\gamma_c)L^{1/\nu}],  
\end{align}
where $a$ is a constant which is set to be $1/2$ in our analysis and $F$ is some smooth function of the single parameter $(\gamma-\gamma_c)L^{1/\nu}$. 
The exponent $\nu$ is related to the divergence of a correlation length $\xi \sim |\gamma-\gamma_c|^{-\nu}$ at the MIC. 
For the critical points $\gamma_c$ estimated from the peak structures of the mutual information in Sec.~\ref{sec:MutualInf}, we perform the scaling analysis for data sets in the range $\gamma \in [\gamma_c-0.5,\gamma_c+0.5]$ and for systems sizes from $L=8$ to 20. 
We obtain the exponents $\nu$ by the Bayesian scaling analysis distributed in Ref.~\cite{Bayesian}, which are summarized in Table~\ref{table:Exponents}. 
The resulting data collapses are shown in Fig.~\ref{fig:ScalingCollapse}.
We have also performed the scaling analysis without using \textit{a priori} estimates of $\gamma_c$ from the mutual information but using the same data sets as detailed in Appendix~\ref{app:Scaling}. 
The obtained values of $\gamma_c$ roughly agree with those estimated from the mutual information within statistical error. 
In both analyses, we find that the exponent $\nu$ takes values from 1.1 to 1.3.

\begin{figure*}
\includegraphics[clip,width=\textwidth]{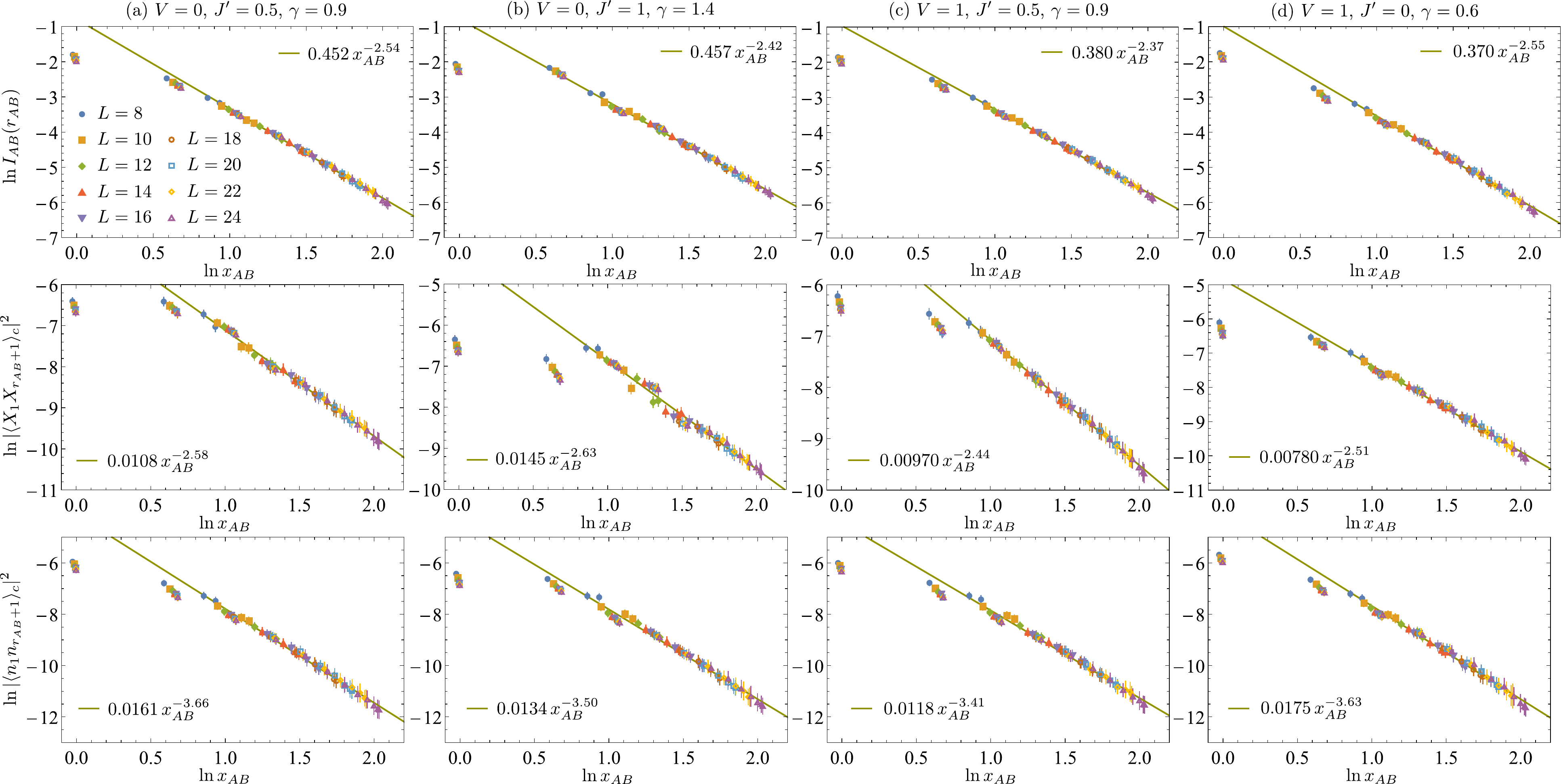}
\caption{Steady-state values of the mutual information and the absolute squares of the connected correlation functions of $X_j$ and $n_j$ at the entanglement transition $\gamma=\gamma_c$ are plotted against the chord distance $x_{AB}$ between two sites. 
The logarithmic scales are used for both axes. 
Each data point is averaged over 400 quantum trajectories. 
The solid lines are fitting functions of the form $D x_{AB}^{-2\Delta}$. 
For the correlation functions, fitting is performed for data points with $\ln x_{AB} \geq 1$.}
\label{fig:MIScaling}
\end{figure*}

We also study the scaling behavior of the steady-state mutual information $I_{AB}(r_{AB})$ as a function of the distance $r_{AB}$ between two sites at the entanglement transition $\gamma=\gamma_c$. 
Similarly to the entanglement entropy, we use the chord distance $x_{AB}=(L/\pi) \sin (\pi r_{AB}/L)$. 
The results are shown in the top panels of Fig.~\ref{fig:MIScaling}.
The mutual informations exhibits power-law behaviors $I_{AB} \propto x_{AB}^{-2\Delta}$ for large distances, which are predicted for 1D CFTs \cite{Furukawa09, Calabrese09, Calabrese11} and also confirmed for the MICs in unitary circuit models \cite{Skinner19, Li19}. 
The corresponding exponents $\Delta$ are shown in Table~\ref{table:Exponents}. 
As opposed to the observation made for the Bose-Hubbard model with random projective measurements \cite{Tang20}, our analysis shows that the mutual information is fitted well into a power-law form with an exponent $\Delta$ independent of the system size, although we can only access relatively small systems compared with Ref.~\cite{Tang20}. 

We also study the scaling behaviors of the absolute squares of the connected correlation functions $\langle X_1 X_{r_{AB}+1} \rangle_c$ and $\langle n_1 n_{r_{AB}+1} \rangle_c$ in the steady-state regimes. which are shown as functions of the chord distance $x_{AB}$ in the bottom two panels of Fig.~\ref{fig:MIScaling}. 
Both correlations functions are fitted well into power-law forms as anticipated for quantum critical systems. 
Furthermore, we found that exponents $\Delta_X$ for $|\langle X_1 X_{r_{AB}+1} \rangle_c|^2 \propto x_{AB}^{-2\Delta_X}$ are close to the exponents $\Delta$ for the mutual informations. 
As the exponent $\Delta$ of the mutual information for small subregions $A$ and $B$ but a large distance between them is the scaling dimension of a leading scaling operator \cite{Calabrese11}, this indicates that the operator $X_j$ is related to such a leading operator giving the most dominant correlation at the entanglement transition. 
This is consistent with the fact that the correlation $|\langle X_1 X_{L/2+1} \rangle_c|^2$ as a function of the measurement strength $\gamma$ shows the similar peak structure to the mutual information as observed in Sec.~\ref{sec:MutualInf}. 
On the other hand, the exponents $\Delta_n$ for $|\langle n_1 n_{r_{AB}+1} \rangle_c|^2 \propto x_{AB}^{-2\Delta_n}$ are generically larger than those for the mutual information and $|\langle X_1 X_{r_{AB}+1} \rangle_c|^2$. 
This could be related to their functional forms in $\gamma$ that do not follow those for the mutual informations. 
The exponents $\Delta_X$ and $\Delta_n$ at the MICs are also summarized in Table~\ref{table:Exponents}.

\subsection{$U(1)$-symmetry resolved entanglement}
\label{sec:SymResEnt}

\begin{figure}
\includegraphics[clip,width=0.48\textwidth]{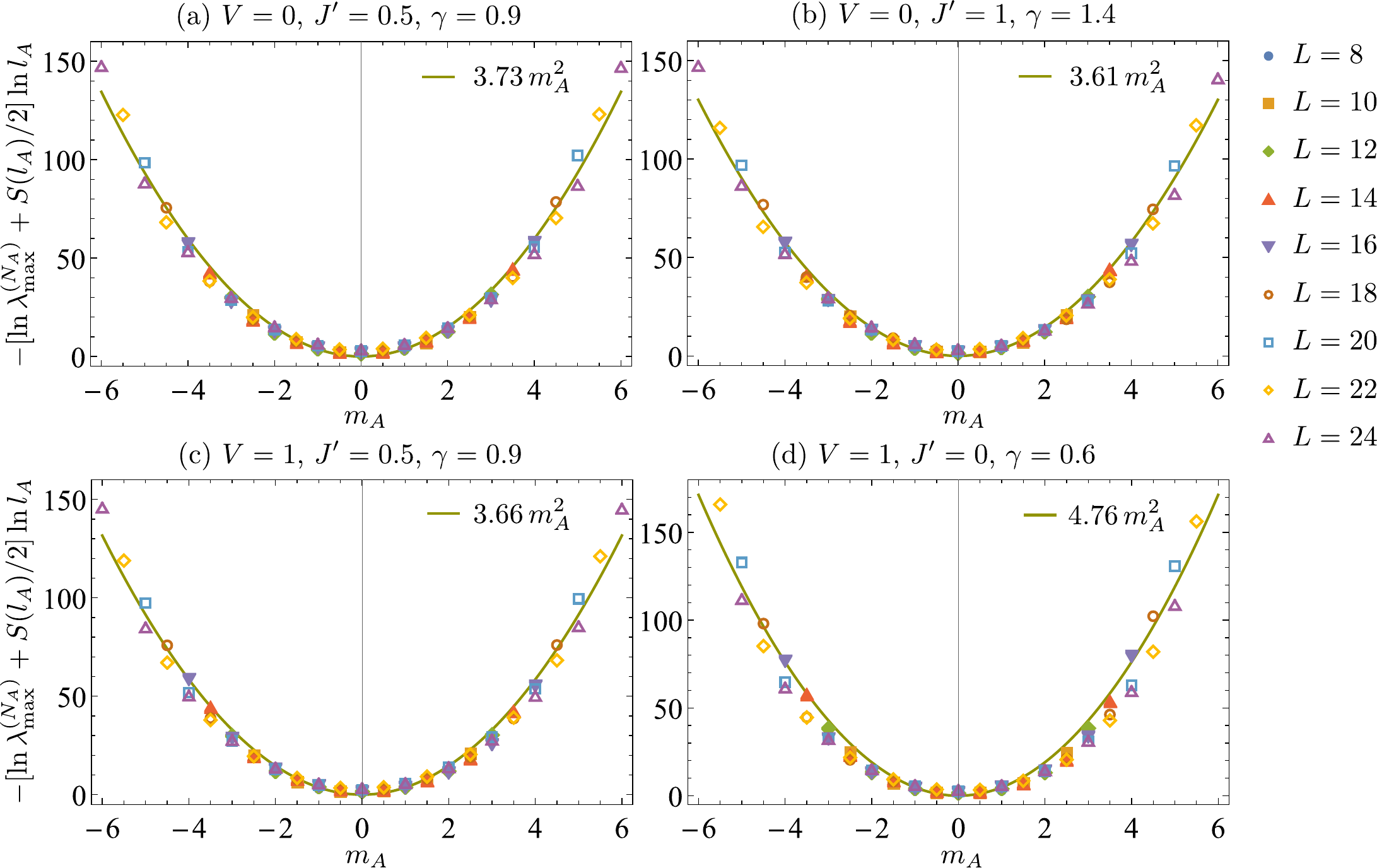}
\caption{Largest eigenvalues of the symmetry-resolved reduced density matrix $\rho_A^{(N_A)}$ for $l_A=L/2$ are shown in the form $-[\ln \lambda_\textrm{max}^{(N_A)}(l_A) +S(l_A)/2] \ln l_A$ as functions of $m_A = N_A -l_A/2$. 
Each data point is taken at $t=100$ for nonintegrable models (a)-(c) and at $t=200$ for an integrable model (d) and then averaged over 400 quantum trajectories. 
The solid lines are fitting functions quadratic in $m_A$.}
\label{fig:SymResESMax}
\end{figure}

We finally examine entanglement measures particular for $U(1)$-symmetry conserving systems, called the $U(1)$-symmetry resolved entanglement entropy \cite{Laflorencie14, Goldstein18, Xavier18}. 
Since the entire quantum trajectory dynamics conserves the total particle number $n_\textrm{tot}$, we have $[\rho(t), n_\textrm{tot}]=0$. 
Tracing out the degrees of freedom in the subsystem $\bar{A}$ in this expression yields $[\rho_A(t), n_A]=0$ where $n_A = \sum_{j \in A} n_j$ is the total particle number operator in the subsystem $A$. 
Thus, the reduced density matrix $\rho_A(t)$ computed from a quantum trajectory takes a block-diagonal form, 
\begin{align}
\rho_A(t) = \bigotimes_{N_A=0}^{l_A} \rho^{(N_A)}_A(t),
\end{align}
where each block $\rho_A^{(N_A)}(t)$ is associated with an eigenvalue $N_A$ of the operator $n_A$.
The block density matrix can be written as 
\begin{align}
\rho_A^{(N_A)}(t) = \mathcal{P}_A^{(N_A)} \rho_A(t) \mathcal{P}_A^{(N_A)},
\end{align}
where $\mathcal{P}_A^{(N_A)}$ is a projection operator onto the subspace of the particle number $N_A$ and can be defined through the spectral decomposition $n_A = \sum_{N_A=0}^{l_A} N_A \mathcal{P}_A^{(N_A)}$. 
We note that $\rho_A^{(N_A)}(t)$ is not normalized, adopting the convention of Ref.~\cite{Goldstein18}. 
We then consider the largest eigenvalue $\lambda^{(N_A)}_\textrm{max}(l_A,t)$ of the block reduced density matrix $\rho^{(N_A)}_A(t)$, which is related to the symmetry-resolved Renyi entanglement entropy in the limit of Renyi index $n \to \infty$. 
We expect that $\lambda^{(N_A)}_\textrm{max}(l_A)$ in a steady-state regime at the MIC is related to the scaling dimension $\Delta(N_A)$ of an operator associated to $N_A$ in a boundary CFT by
\begin{align} \label{eq:EntSpecScaling}
-\ln \lambda^{(N_A)}_\textrm{max}(l_A) = \frac{S(l_A)}{2} + C \frac{\Delta(N_A)}{\ln l_A}.
\end{align}
Here, the first term comes from the lowest entanglement spectrum $-\ln \lambda_\textrm{max}(l_A)$ of the full reduced density matrix $\rho_A(t)$, which is called the single-copy entanglement. 
In the (1+1)-D CFT, the scaling form of the single-copy entanglement has been known to be half of the von Neumann entanglement entropy $S(l_A)/2$ \cite{Eisert05, Orus06}. 
We adopt $S(l_A)/2$ in the scaling form \eqref{eq:EntSpecScaling} instead of using $-\ln \lambda_\textrm{max}(l_A)$ directly, since $\lambda_\textrm{max}(l_A)$ is almost degenerate between $N_A=(L/2 \pm 1)/2$ sectors for system sizes $L \in 4 \mathbb{Z}+2$ and the subsystem size $l_A=L/2$. 
The second term in Eq.~\eqref{eq:EntSpecScaling} is expected from the scaling form of entanglement spectral gaps in 1D critical systems, which is inversely proportional to $\ln l_A$ and predicted by a boundary CFT \cite{Lauchli13, Ohmori15, Cardy16}.
In Fig.~\ref{fig:SymResESMax}, we show $-\ln \lambda_\textrm{max}^{(N_A)}(l_A)$ subtracted by $S(l_A)/2$ under the bipartition of the system $l_A=L/2$ as functions of the particle number deviation from the mean value $m_A = N_A -l_A/2$ in steady states.
The scaling dimension $\Delta (m_A)$ extracted from $\lambda_\textrm{max}^{(N_A)}(l_A)$ up to a numerical constant $C$ are fitted into quadratic functions $C'm_A^2$, which are reminiscent of those for a free boson CFT or a Tomonaga-Luttinger liquid theory that describes conventional critical systems preserving $U(1)$ symmetry \cite{Goldstein18}. 
Also, the coefficients $C'$ for nonintegrable models [Fig.~\ref{fig:SymResESMax} (a)--(c)] take similar values. 
This may indicate that $C'$ is a universal value related to the exponents $\nu$ or $\Delta$ obtained in Sec.~\ref{sec:Scaling}, which also take similar values. 
In the Tomonaga-Luttinger liquid theory, the coefficient $C'$ is proven to be a universal quantity depending only on the scaling dimension of a leading operator \cite{Goldstein18}. 
Thus, the underlying CFT of our MICs with $U(1)$ symmetry may have an operator content similar to the Tomonaga-Luttinger liquid theory in conventional critical systems. 
However, we note that in the Tomonaga-Luttinger liquid theory, the mutual information takes a functional form $I_{AB}(r_{AB}) \propto x_{AB}^{-2\min(\eta,1/\eta)}$ with a universal constant (or the Luttinger parameter) $\eta$ and thus the exponent $\Delta$ cannot exceed 1 \cite{Furukawa09, Calabrese09, Calabrese11}; this is not the case for the MICs as observed in Sec.~\ref{sec:Scaling}. 
Therefore, the observed MICs should belong to different universality classes from the standard Tomonaga-Luttinger liquid theory even in the presence of $U(1)$ symmetry.

\section{Applications to experiments}
\label{sec:ApplExp}

In contrast to quantum circuit models, our theoretical consideration is directly relevant to actual physical systems in, e.g., ultracold gases.
Specifically, to realize the model~\eqref{eq:XXZHam}, one can use the ladder-type optical lattice illustrated in Fig.~\ref{fig:schem}. 
A concrete protocol to realize such a zigzag lattice by the current techniques can be found in Ref.~\cite{ZT15}; a similar ladder structure has already been realized in a number of experiments (see, e.g., Ref.~\cite{AM14}). 
A hard-core bosonic gas can naturally be realized by working in the Tonks-Girardeau regime via, for instance, employing tight light-induced confinements \cite{PB04,KT04}.  
The system can be prepared in the N\'{e}el state \cite{Trotzky12, Schreiber15} as assumed in the present numerical simulations, while we expect that a specific choice of the initial state is irrelevant to steady-state properties as long as the filling is the same.
Finally, the continuous measurement process corresponding to the jump operator $L_j=n_j$ can be realized by combining dispersive light scattering with the high-resolution optical setup as realized in quantum gas microscopy \cite{Bakr09}.

A quantity that we are primarily interested in at the MIC is the entanglement entropy. 
While the von Neumann entanglement entropy mainly analyzed in this study is rather challenging to access in the currently available techniques, the second Renyi entanglement entropy $S_2 = -\ln \textrm{Tr}_A (\rho_A^2)$ has experimentally been measured from quantum interference between two identical copies of quantum many-body states \cite{Islam15} or by randomized measurements \cite{Brydges19}. 
We note that there might be a Renyi index dependence for the critical measurement strength and universal quantities at the MIC, such as $\alpha_S$ and $\nu$, as observed in Refs.~\cite{Skinner19, Li19, Zabalo20}. 
Thus, critical properties obtained from the second Renyi entropy can in principle be different from those obtained from the von Neumann entropy, which have been studied in this paper. 

\begin{figure}
\includegraphics[clip,width=0.4\textwidth]{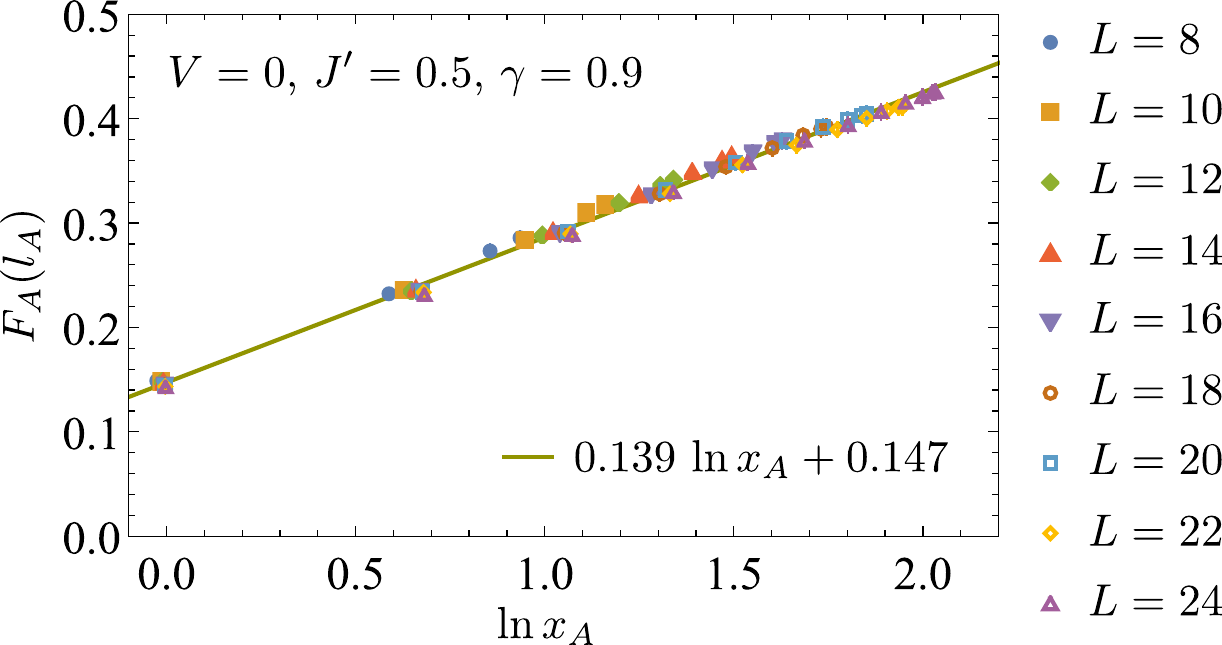}
\caption{Bipartite particle number fluctuation $F_A(l_A)$ plotted against the logarithm of the chord distance $x_A$ of the subsystem $A$ for $L=8$ to $24$. 
The data are shown for a nonintegrable model with $V=0$ and $J'=0.5$ at the critical measurement strength $\gamma=0.9$. 
Each data point is the averaged value over 400 quantum trajectories and over the time interval $t \in [50,100]$. 
The solid line is the fitting function of the form $\alpha_F \ln x_A +\beta_F$.}
\label{fig:BP}
\end{figure}

Meanwhile, since our model preserves the total particle number, one could instead focus on another useful quantity called the bipartite particle-number fluctuation \cite{Klich09, Song10, Song11, Song12}, 
\begin{align}
F_A(l_A) = \langle (n_A -\langle n_A \rangle)^2 \rangle, 
\end{align}
where $n_A$ is the total particle number operator in the subsystem $A$ as given above: $n_A = \sum_{j \in A} n_j$. 
The bipartite fluctuation $F_A(l_A)$ is relatively easy to measure compared to the entanglement entropy itself in quantum gas microscopy or quantum point contact  \cite{Klich09, Song12}.
This quantity shares similar properties with the von Neumann entanglement entropy $S(l_A)$ in many aspects; in particular, it exhibits a logarithmic scaling in 1D critical systems described by the Tomonaga-Luttinger liquid theory \cite{Song10}. 
In Fig.~\ref{fig:BP}, we show the steady-state values of the bipartite fluctuation $F_A$ computed for quantum trajectories at the MIC for a nonintegrable model with $V=0$ and $J'=0.5$.
It clearly exhibits a logarithmic scaling of the form $\alpha_F \ln x_A +\beta_F$ with respect to the chord length $x_A$ of the subsystem $A$, as expected.\footnote{
The Tomonaga-Luttinger liquid theory predicts that the coefficient $\alpha_F$ for this logarithmic scaling of $F_A(l_A)$ and the coefficient $C'$ for the quadratic scaling of the $U(1)$-symmetry resolved entanglement in Eq.~\eqref{eq:EntSpecScaling} are both related to the Luttinger parameter $K=1/2\eta$ by $\alpha_F = K/\pi^2$ and $C' = \pi^2/2K$ \cite{Song10, Goldstein18}.
We thus find $\alpha_F = 1/2C'$. 
From the result of Sec.~\ref{sec:SymResEnt}, we obtain $C' \sim 3.73$ for the same model and then $1/2C' \sim 0.134$. 
Therefore, this relation also holds for our MIC as $\alpha_F \sim 0.139$, providing another evidence supporting a Tomonaga-Luttinger-liquid-like feature of the MIC with the $U(1)$ symmetry.
}

\begin{figure}
\includegraphics[clip,width=0.4\textwidth]{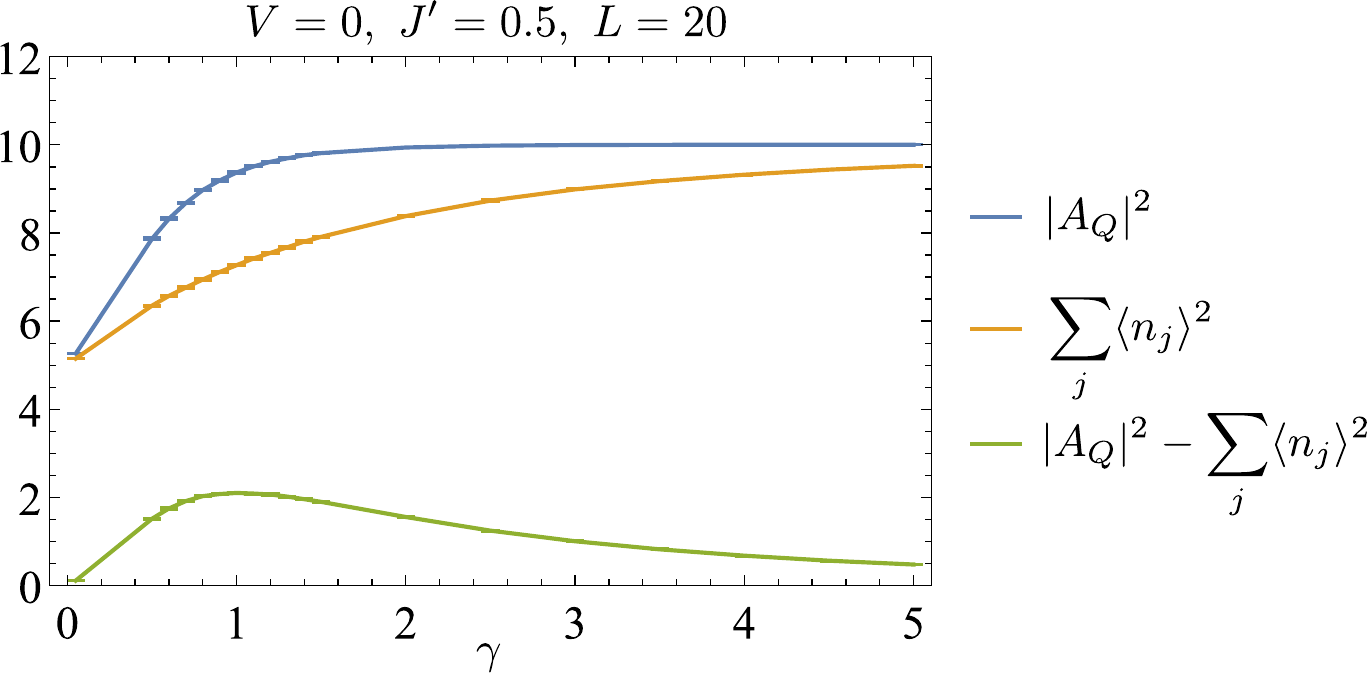}
\caption{Absolute square of the interference amplitude $\langle |A_Q|^2 \rangle$ defined in Eq.~\eqref{eq:AQ} is plotted against the measurement strength $\gamma$ for the $L=20$ system of a nonintegrable model with $V=0$ and $J'=0.5$. 
When the $j=k$ contribution, $\sum_j \langle n_j \rangle^2$, is subtracted, $\langle |A_Q|^2 \rangle$ shows a peak structure.
Each data point is the averaged value over 800 quantum trajectories and over the time interval $t \in [50,100]$.}
\label{fig:AQ}
\end{figure}

As discussed in Sec.~\ref{sec:Results}, locating the MIC only by the logarithmic scalings of the entanglement entropy or bipartite fluctuation might not be an easy task for small-size systems. 
One can then use peak structures of the mutual information or correlation functions to pin down the entanglement transition. 
As discussed above, the mutual information defined through the second Renyi entropy can in principle be measured in experiment \cite{Islam15}. 
While the correlation functions of particle number operators $n_j$ would be easiest to measure, they do not show peak structures at the MIC. 
On the other hand, the correlation functions of $X_j = (b_j^\dagger +b_j)/2$ instead show peaks and might be  measured by combining Ramsey spectroscopy with the state-selective, site-resolved projection measurement \cite{Hazzard14, FT15}. 
The other quantity, which is perhaps much more accessible in ultracold atomic measurements, is related to the complex amplitude of a two-point operator between two identical but independent systems, summed over the system length, 
\begin{align}
A_Q(L) = \sum_{j=1}^L b^\dagger_{1,j} b_{2,j},
\end{align}
where $b_{1,j}$ and $b_{2,j}$ are boson annihilation operators in the independent systems. 
The expectation value of this quantity itself becomes zero, but that of its absolute square, 
\begin{align} \label{eq:AQ}
\langle |A_Q(L)|^2 \rangle = \sum_{j=1}^L \sum_{k=1}^L \langle b^\dagger_{1,j} b_{1,k} \rangle \langle b^\dagger_{2,k} b_{2,j} \rangle, 
\end{align}
is a meaningful quantity, which can be extracted as the average contrast in the time-of-flight interference measurement \cite{Polkovnikov06, Hofferberth08}. 
Since the two systems are identical, this can be computed from two-point correlation functions of a single system. 
The steady-state values of $\langle |A_Q(L)|^2 \rangle$ are shown in Fig.~\ref{fig:AQ} for the $L=20$ system of a nonintegrable model with $V=0$ and $J'=0.5$.
We find that $\langle |A_Q(L)|^2 \rangle$ monotonically increases as the measurement strength $\gamma$ is increased. 
This behavior comes from the $j=k$ contribution in the summation of Eq.~\eqref{eq:AQ}, which is the sum of the squared expectation values of the occupation numbers, $\sum_j \langle n_j \rangle^2$. 
This quantity goes to $L/4$ in the limit $\gamma \to 0$ as $\langle n_j \rangle$ relaxes to $\nu=1/2$, and it monotonically grows and reaches $L/2$ in the limit $\gamma \to \infty$ as the quantum trajectory is frozen to a product state by frequent measurements.
Subtracting this contribution from $\langle |A_Q(L)|^2 \rangle$, we find a peak around the critical measurement strength $\gamma=0.9$. 
Thus, $\langle |A_Q(L)|^2 \rangle$ could also be useful to locate the MIC.

\section{Conclusion}
\label{sec:Conclusion}

We studied the quantum trajectory dynamics of many-body quantum entanglement under continuous measurement. 
We considered a hard-core boson chain subject to continuous measurement of local occupation numbers in both nonintegrable and integrable regimes. 
As observed for circuit models, we find the measurement-induced entanglement transition from the volume-law to area-law regimes of the entanglement as the measurement strength is increased. 
The transition point can be located by peak structures of the mutual information or correlation functions of the boson operators. 
At the transition, the von Neumann entanglement entropy scales logarithmically in the subsystem size and the mutual information and correlation functions decay algebraically, indicating emergent conformal invariance and thus the MIC. 
We extract various critical exponents at the MIC, which appear not to fit into any known universality class of the conventional critical systems described by the (1+1)-D CFT with the $U(1)$ symmetry or the MICs in circuit models. 
However, we also observe some features reminiscent of a free boson CFT or a Tomonaga-Luttinger liquid theory, such as quadratic scaling in the $U(1)$-symmetry resolved entanglement and logarithmic scaling in the bipartite particle-number fluctuation.

We emphasize that the entanglement transition is a phenomenon beyond the unconditional dynamics obeying the Lindblad master equation, where the system in general monotonically heats up to the mixed-state density matrix of an infinite-temperature equilibrium state and thus the steady-state properties become featureless.
Due to the $U(1)$ symmetry and quick relaxation of the local particle densities under the time evolution, the quantum trajectory dynamics in our model is close to that for unitary circuit models with random projective measurements, except that the unitary dynamics is effectively given by the unitary time evolution by the Hamiltonian. 
However, the trajectory dynamics can be more complex in the absence of $U(1)$ symmetry, since the time evolution between projective measurements (quantum jumps) is in general the nonunitary dynamics given by a nontrivial non-Hermitian Hamiltonian and may result in inhomogeneity in the waiting-time distribution for quantum jumps and the probability distribution of measured sites. 
Therefore, it is interesting to ask how the occasion for the measurement-induced transition is extended to more general setups of the quantum trajectory dynamics, such as the one under a global measurement \cite{Ivanov20, Kroeger20}.

Another question is the relation between integrability of the Hamiltonian and the measurement-induced entanglement transition. 
While we found the transition even in an integrable regime of the model, it has been argued that for free-fermion chains the volume-law entanglement regime is immediately lost for any finite strength of the measurement \cite{Cao19}. 
The latter scenario stems from the collapsed quasiparticle pair ansatz, which is based on the quasiparticle pair ansatz proposed to explain the linear growth of entanglement in the quench dynamics and the resulting volume-law entanglement in steady states for integrable models \cite{Calabrese05, Calabrese16}. 
Thus, the collapsed quasiparticle pair ansatz could also be applied to general integrable models, implying the absence of entanglement transition.
Since our numerical simulation is limited to small-size systems, it is possible that the scaling ansatz fails for large systems and thus the entanglement transition disappears. 
However, as the authors of Ref.~\cite{Cao19} noted, there can be more complicated dynamics beyond the quasiparticle pair ansatz, which makes the entanglement transition feasible. 
In fact, Alberton \textit{et al.} have recently suggested that even for free fermion chains there can be an entanglement phase transition, depending on specific ways of unraveling the Lindblad master equation to define quantum trajectories \cite{Alberton20}. 
They found that both the quantum jump process used in our study and a quantum diffusion (Wiener) process used in Ref.~\cite{Cao19} lead to transitions from the subvolume-law (logarithmic) to area-law entanglement phase when the measurement strength is increased. 
This raises another possibility that integrable systems do not actually possess a volume-law entanglement phase but a subvolume-law phase, whose scaling behavior can be accessed only for sufficiently large system sizes. 
We believe that our study stimulates more detailed studies about scaling properties of the entanglement and their trajectory dependence on the measurement-induced dynamics of many-body quantum systems.

Last but not least, our model is directly relevant for experimental realizations of the entanglement transition in ultracold gases, where the nonunitary dynamics subject to site-resolved continuous position measurement can be implemented via quantum gas microscopy. 
Apart from direct measurement of the entanglement entropy, we propose measuring the bipartite fluctuation of the particle number as a useful quantity to extract the logarithmic scaling behavior at the MIC, thanks to the total particle number conservation. 
The absolute square of the interference amplitude, $\langle |A_Q|^2 \rangle$, can also be used to pin down the location of the MIC when the measurement strength is controlled. 
We hope that our study stimulates further studies and offers a foothold for experimentally exploring nontrivial entanglement dynamics in open nonequilibrium many-body systems. 

\acknowledgments
Y.F. is not supported by any funding, but he is grateful to Condensed Matter Theory Laboratory in RIKEN for their generosity, which allows him to use their computer facilities. 
Y.A. is grateful to Takeshi Fukuhara and Takahiro Sagawa for useful discussions. 
Y.A. acknowledges support from the Japan Society for the Promotion of Science through Grant No. JP19K23424.

\appendix
\section{Scaling collapse of entanglement curves}
\label{app:Scaling}

In Sec.~\ref{sec:Scaling}, we performed data collapses of the steady-state entanglement entropy into the scaling form \eqref{eq:ScalingEE} with critical measurement strengths $\gamma_c$ \textit{a priori} estimated from the peak structures of the mutual information in Sec.~\ref{sec:MutualInf}. 
However, we can directly work on the scaling form \eqref{eq:ScalingEE} to estimate both $\gamma_c$ and the correlation length exponent $\nu$ without prior knowledge about them. 
To do so, we employ an algorithm proposed in Ref.~\cite{Skinner19} to perform this scaling analysis. 
Namely, one can determine a set of curves,
\begin{align}
y_L = S(\gamma, L; l_A) -S(\gamma_c, L; l_A),
\end{align}
for a given data set and a given value of $\gamma_c$ and $\nu$, where $S(\gamma,L; l_A)$ with $\gamma$ not included in the data set is obtained by a linear interpolation. 
The curve $y_L$ for a given $L$ is a function of $\gamma$ and thereby a function of the scaling variable $x=(\gamma-\gamma_c)L^{1/\nu}$. 
One can then define an objective function as the sum of the mean-square deviations of each $y_L(x)$ from their common mean, 
\begin{align}
R = \sum_{i,L} [y_L(x_i) -\bar{y}(x_i)]^2,
\end{align}
where $x_i$ is a scaling variable obtained from the data set and $\bar{y}(x)$ is the average of $y_L(x)$ over different system sizes $L$. 
The value of $y_L(x)$ with $x$ not included in the data set is again determined by a linear interpolation. 
An optimal estimate of $\gamma_c$ and $\nu$ will be obtained by minimizing the objective function $R(\gamma_c,\nu)$.

\begin{figure}
\includegraphics[clip,width=0.48\textwidth]{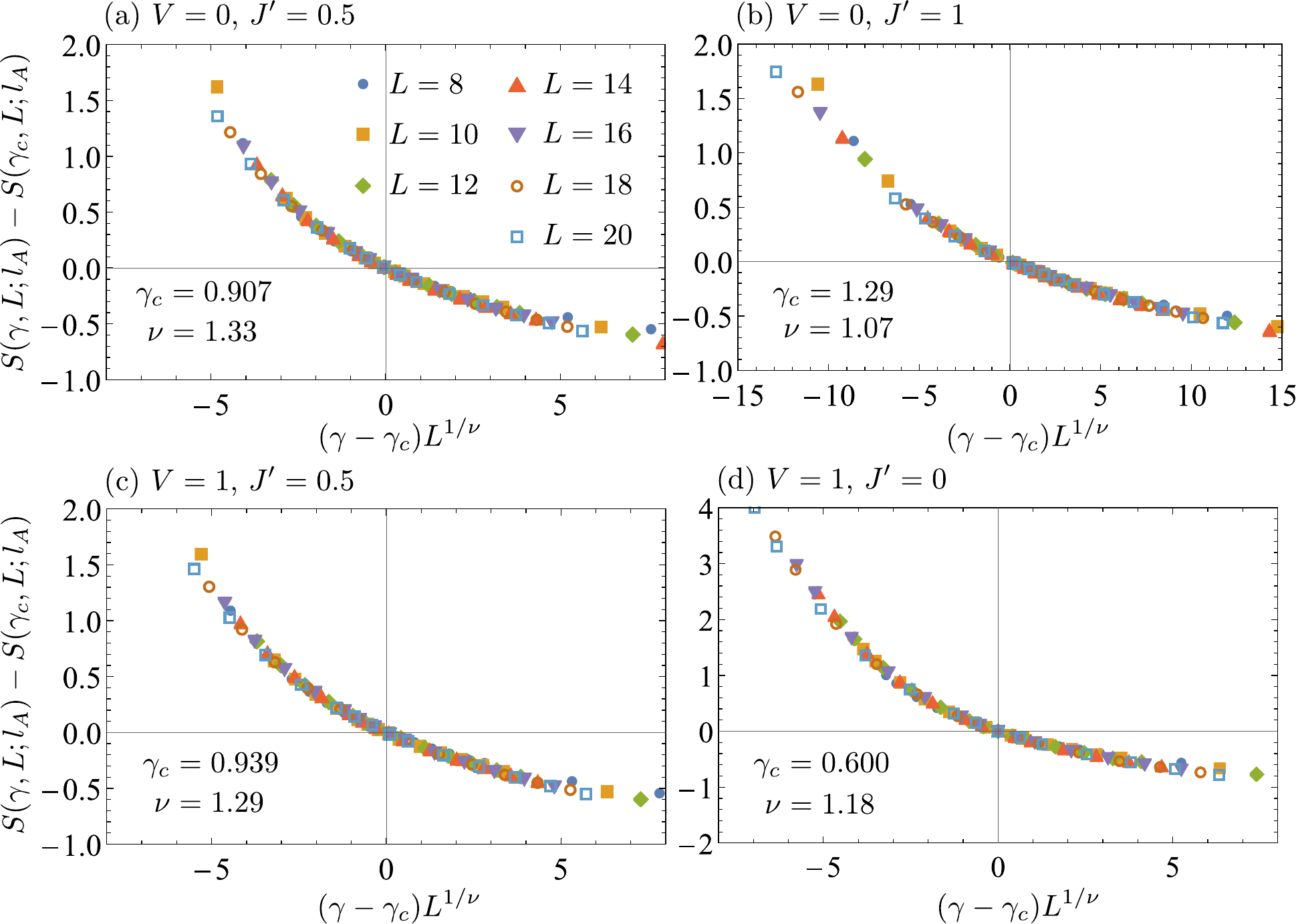}
\caption{Data collapses of the steady-state entanglement entropy into the scaling form \eqref{eq:ScalingEE}, obtained by the search algorithm presented in Ref.~\cite{Skinner19}. 
The data are shown for nonintegrable models (a)--(c) and an integrable model (d). 
Each data point is averaged over 800 quantum trajectories.}
\label{fig:ScalingCollapseS}
\end{figure}

We applied the above algorithm to search for the best estimates of $\gamma_c$ and $\nu$ and found that estimated values of $\gamma$ substantially deviate to those in presumed volume-law regimes when the data sets for all values of $\gamma \in [0.05,5]$ are used. 
We thus use the same data sets used in Sec.~\ref{sec:Scaling} to estimate the exponents $\nu$: $\gamma \in [0.4,1.4]$ for $(V,J')=(0,0.5)$ and $(1,0.5)$, $\gamma \in [0.9,1.9]$ for $(V,J')=(0,1)$, and $\gamma \in [0.1,1.1]$ for $(V,J')=(1,0)$. 
The obtained data collapses are shown in Fig.~\ref{fig:ScalingCollapseS} and the values of $\gamma_c$ and $\nu$ are given in Table~\ref{table:Exponents2}, where the comparison with the results in the main text is also made. 
\begin{table*}
\begin{ruledtabular}
\begin{tabular}{lcccc}
Model & $(V,J')=(0,0.5)$ & $(V,J')=(0,1)$ & $(V,J')=(1,0.5)$ & $(V,J')=(1,0)$ \\
\hline
$\gamma_c$ & 0.907 & 1.29 & 0.939 & 0.600 \\
$\nu$ & 1.33 & 1.07 & 1.29 & 1.18 \\
\hline
$\gamma_c$ in Sec.~\ref{sec:MutualInf} & 0.9 & 1.4 & 0.9 & 0.6 \\
$\nu$ in Sec.~\ref{sec:Scaling} & 1.34(3) & 1.22(3) & 1.26(3) & 1.21(2)
\end{tabular}
\end{ruledtabular}
\caption{Estimates of the critical measurement strength $\gamma_c$ and the exponents $\nu$ obtained by the search algorithm presented in Ref.~\cite{Skinner19}. 
The results are compared with those obtained in Secs.~\ref{sec:MutualInf} and \ref{sec:Scaling}.}
\label{table:Exponents2}
\end{table*}
The obtained values of $\gamma_c$ roughly agree with those obtained in Sec.~\ref{sec:MutualInf} as they reside in ranges of the broad peaks of the mutual information within statistical error, as shown in Fig.~\ref{fig:MutualInfSS} for larger system sizes. 

\bibliography{MICContMonit}

\end{document}